\documentclass[usenatbib,a4paper]{mn2e}
\usepackage{amsmath}
\usepackage{amssymb}
\usepackage{amsfonts}
\usepackage {multirow}
\usepackage {graphicx}
\usepackage {natbib}
\usepackage{txfonts}
\usepackage{graphicx}
\usepackage{soul}
\usepackage{float}
\usepackage[hyperindex,breaklinks=true, colorlinks, citecolor=blue]{hyperref}
\usepackage{tablefootnote}
\usepackage{eurosym}
\usepackage{ulem}
\usepackage[usenames,dvipsnames]{xcolor}
\usepackage{booktabs}
\usepackage[mathscr]{euscript}
\usepackage{multicol}
\usepackage{multirow}
\usepackage{makecell}
\usepackage{pifont}
\usepackage{slashed}
\usepackage{verbatim}
\usepackage{aas_macros}
\usepackage{ae,aecompl}
\usepackage{mathtools}
\usepackage[english]{babel}
\usepackage{float} 
\usepackage{color}
\usepackage{lscape}

\DeclareSymbolFont{rsfs}{U}{rsfs}{m}{n}
\DeclareSymbolFontAlphabet{\mathscrsfs}{rsfs}

\def\setsymbol#1#2{\expandafter\def\csname #1\endcsname{#2}}
\def\getsymbol#1{\csname #1\endcsname}

\newbox\tablebox    \newdimen\tablewidth
\def\leaderfil{\leaders\hbox to 5pt{\hss.\hss}\hfil}
\def\tablenote#1 #2\par{\begingroup \parindent=0.8em
    \abovedisplayshortskip=0pt\belowdisplayshortskip=0pt
    \noindent
    $$\hss\vbox{\hsize\tablewidth \hangindent=\parindent \hangafter=1 \noindent
    \hbox to \parindent{$^#1$\hss}\strut#2\strut\par}\hss$$
    \endgroup}

\def\L2{\ifmmode L_2\else $L_2$\fi}

\def\DeltaT{\ifmmode \Delta T\else $\Delta T$\fi}
\def\deltat{\ifmmode \Delta t\else $\Delta t$\fi}
\def\fknee{\ifmmode f_{\rm knee}\else $f_{\rm knee}$\fi}
\def\Fmax{\ifmmode F_{\rm max}\else $F_{\rm max}$\fi}
\def\solar{\ifmmode{\rm M}_{\mathord\odot}\else${\rm M}_{\mathord\odot}$\fi}
\def\Msolar{\ifmmode{\rm M}_{\mathord\odot}\else${\rm M}_{\mathord\odot}$\fi}
\def\Lsolar{\ifmmode{\rm L}_{\mathord\odot}\else${\rm L}_{\mathord\odot}$\fi}
\def\inv{\ifmmode^{-1}\else$^{-1}$\fi}
\def\mo{\ifmmode^{-1}\else$^{-1}$\fi}
\def\sup#1{\ifmmode ^{\rm #1}\else $^{\rm #1}$\fi}
\def\expo#1{\ifmmode \times 10^{#1}\else $\times 10^{#1}$\fi}
\def\,{\thinspace}
\def\lsim{\mathrel{\raise .4ex\hbox{\rlap{$<$}\lower 1.2ex\hbox{$\sim$}}}}
\def\gsim{\mathrel{\raise .4ex\hbox{\rlap{$>$}\lower 1.2ex\hbox{$\sim$}}}}

\def\simprop{\mathrel{\raise .4ex\hbox{\rlap{$\propto$}\lower 1.2ex\hbox{$\sim$}}}}
\def\deg{\ifmmode^\circ\else$^\circ$\fi}
\def\pdeg{\ifmmode $\setbox0=\hbox{$^{\circ}$}\rlap{\hskip.11\wd0 .}$^{\circ}
          \else \setbox0=\hbox{$^{\circ}$}\rlap{\hskip.11\wd0 .}$^{\circ}$\fi}
\def\arcs{\ifmmode {^{\scriptstyle\prime\prime}}
          \else $^{\scriptstyle\prime\prime}$\fi}
\def\arcm{\ifmmode {^{\scriptstyle\prime}}
          \else $^{\scriptstyle\prime}$\fi}
\newdimen\sa  \newdimen\sb
\def\parcs{\sa=.07em \sb=.03em
     \ifmmode \hbox{\rlap{.}}^{\scriptstyle\prime\kern -\sb\prime}\hbox{\kern -\sa}
     \else \rlap{.}$^{\scriptstyle\prime\kern -\sb\prime}$\kern -\sa\fi}
\def\parcm{\sa=.08em \sb=.03em
     \ifmmode \hbox{\rlap{.}\kern\sa}^{\scriptstyle\prime}\hbox{\kern-\sb}
     \else \rlap{.}\kern\sa$^{\scriptstyle\prime}$\kern-\sb\fi}
\def\ra[#1 #2 #3.#4]{#1\sup{h}#2\sup{m}#3\sup{s}\llap.#4}
\def\dec[#1 #2 #3.#4]{#1\deg#2\arcm#3\arcs\llap.#4}
\def\deco[#1 #2 #3]{#1\deg#2\arcm#3\arcs}
\def\rra[#1 #2]{#1\sup{h}#2\sup{m}}

\def\dots{\relax\ifmmode \ldots\else $\ldots$\fi}
\def\WHzsr{\ifmmode $W\,Hz\mo\,sr\mo$\else W\,Hz\mo\,sr\mo\fi}
\def\mHz{\ifmmode $\,mHz$\else \,mHz\fi}
\def\GHz{\ifmmode $\,GHz$\else \,GHz\fi}
\def\mKs{\ifmmode $\,mK\,s$^{1/2}\else \,mK\,s$^{1/2}$\fi}
\def\muKs{\ifmmode \,\mu$K\,s$^{1/2}\else \,$\mu$K\,s$^{1/2}$\fi}
\def\muKRJs{\ifmmode \,\mu$K$_{\rm RJ}$\,s$^{1/2}\else \,$\mu$K$_{\rm RJ}$\,s$^{1/2}$\fi}
\def\muKHz{\ifmmode \,\mu$K\,Hz$^{-1/2}\else \,$\mu$K\,Hz$^{-1/2}$\fi}
\def\MJysr{\ifmmode \,$MJy\,sr\mo$\else \,MJy\,sr\mo\fi}
\def\MJysrmK{\ifmmode \,$MJy\,sr\mo$\,mK$_{\rm CMB}\mo\else \,MJy\,sr\mo\,mK$_{\rm CMB}\mo$\fi}
\def\microns{\ifmmode \,\mu$m$\else \,$\mu$m\fi}

\def\muK{\ifmmode \,\mu$K$\else \,$\mu$\hbox{K}\fi}
\def\microK{\ifmmode \,\mu$K$\else \,$\mu$\hbox{K}\fi}
\def\muW{\ifmmode \,\mu$W$\else \,$\mu$\hbox{W}\fi}
\def\kms{\ifmmode $\,km\,s$^{-1}\else \,km\,s$^{-1}$\fi}
\def\kmsMpc{\ifmmode $\,\kms\,Mpc\mo$\else \,\kms\,Mpc\mo\fi}
\providecommand{\sorthelp}[1]{}

\usepackage{xspace}
\newcommand{\xpol}{\ensuremath{\tt Xpol}}
\newcommand{\planck}{\textit{Planck}\xspace}

\newcommand{\wmap}{\textit{WMAP}\xspace}

\DeclareMathAlphabet{\mathsc}{OT1}{cmr}{m}{sc}
\newcommand{\healpix}{\ensuremath{\tt HEALPix}}
\newcommand{\NILC}{\ensuremath{\tt NILC}}
\newcommand{\COMMANDER}{\ensuremath{\tt Commander}}
\newcommand{\SILC}{\ensuremath{\tt SILC}}
\newcommand{\SEVEM}{\ensuremath{\tt SEVEM}}
\newcommand{\LGMCA}{\ensuremath{\tt L-GMCA}}

\newcommand{\GNILC}{\ensuremath{\tt GNILC}}

\newcommand{\smica}{\ensuremath{\tt SMICA}}

\newcommand{\GAL}{{GAL78}}

\newcommand{\Nside}{\ensuremath{N_{\rm side}}}
\newcommand{\dlxx}{\ensuremath{{\cal D}_{\ell}^{XX}}}
\newcommand{\mpfit}{\ensuremath{\tt MPFIT}}
\def\parcm{$^{\scriptstyle\prime}$}

\def\GHz{\ifmmode $\,GHz$\else \,GHz\fi}
\def\muK{\ifmmode \,\mu$K$\else \,$\mu$\hbox{K}\fi}
\def\MJysrmK{\ifmmode \,$MJy\,sr\mo$\,mK$_{\rm CMB}\mo\else \,MJy\,sr\mo\,mK$_{\rm CMB}\mo$\fi}

\def\Em{{$E$}}
\def\Bm{{$B$ }}
\def\Qm{{$Q$ }}
\def\Um{{$U$ }}
\def\bw{\boldsymbol{w }}
\def\ba{{f_{cmb} }}
\def\reff@jnl#1{{\rm#1\/}}

\def\solphys{\reff@jnl{Solar~Phys.}}    
\def\sovast{\reff@jnl{Soviet~Ast.}}     
\def\ssr{\reff@jnl{Space~Sci.Rev.}}    
\def\zap{\reff@jnl{ZAp}}                
\def\nat{\reff@jnl{Nature}}             
\def\aj{\reff@jnl{AJ}}                  
\def\araa{\reff@jnl{ARA\&A}}            
\def\apj{\reff@jnl{ApJ}}                
\def\aapr{\reff@jnl{A\&A~Rev.}}         
\def\aaps{\reff@jnl{A\&AS}}             
\def\azh{\reff@jnl{AZh}}                        
\def\baas{\reff@jnl{BAAS}}              
\def\jcap{\reff@jnl{JCAP}}              
\def\jrasc{\reff@jnl{JRASC}}            
\def\memras{\reff@jnl{MmRAS}}           
\def\mnras{\reff@jnl{MNRAS}}            
\def\pra{\reff@jnl{Phys.Rev.A}}         
\def\prb{\reff@jnl{Phys.Rev.B}}         
\def\prc{\reff@jnl{Phys.Rev.C}}         
\def\prd{\reff@jnl{Phys.Rev.D}}         
\def\prl{\reff@jnl{Phys.Rev.Lett}}      
\def\pasp{\reff@jnl{PASP}}              
\def\pasj{\reff@jnl{PASJ}}              
\def\qjras{\reff@jnl{QJRAS}}            
\def\skytel{\reff@jnl{S\&T}}            
\def\procspie{\reff@jnl{Proceedings of the SPIE}}             
\def\apjl{\reff@jnl{ApJ}}               
\def\apjs{\reff@jnl{ApJS}}              
\def\ao{\reff@jnl{Appl.Optics}}         
\def\apss{\reff@jnl{Ap\&SS}}            
\def\aap{\reff@jnl{A\&A}}   
\title[Dust-Synchrotron template reconstruction using cMILC]
{A new approach of estimating the Galactic thermal dust and synchrotron polarized emission template in the microwave bands}
\author[Debabrata Adak]{Debabrata Adak \thanks{E-mail:~\href{debabrata@iucaa.in}{\textcolor{black}{debabrata@iucaa.in}}}
\\
Inter University Centre for Astronomy and Astrophysics, Post Bag 4, Ganeshkhind, Pune-411007, India\\
}
\date{January 2021}

\begin{document}
\date{\vspace{-6mm}{Accepted  --. Received }}
\maketitle
\begin{abstract}
    The \textit{Internal Linear Combination} (ILC) method has been extensively used to extract the cosmic microwave background (CMB) anisotropy map from foreground contaminated multi-frequency maps. However, the performance of simple ILC is limited and can be significantly improved by heavily constraint equations, dubbed cILC. The standard ILC and cILC works on the spin-0 field. Recently, a generalized version of ILC is developed to estimate polarization maps in which the quantity $Q \pm iU$ is combined at multiple frequencies using complex coefficients called Polarization ILC (PILC). A statistical moment expansion method has recently been developed for high precision modelling of the Galactic foregrounds. This paper develops a semi-blind component separation method combining the moment approach of foreground modelling with a generalized version of the PILC method for heavily constraint equations. The algorithm is developed in pixel space and performs for a spin-2 field. We employ this component separation technique in simultaneous estimation of Stokes $Q$, $U$ maps of the thermal dust at 353 \GHz\ and synchrotron at 30 \GHz\ over 78\% of the sky. We demonstrate the performance of the method on three sets of absolutely calibrated simulated maps at \wmap\ and \planck\ frequencies with varying foreground models.       
\end{abstract}


\section{Introduction}
\label{sec:intro}
Wilkinson Microwave Anisotropy Probe (\wmap, \cite{Bennett:2013}) observed the microwave sky in five frequency bands ranging from 23 to 91 \GHz\ at a resolution which varies between 52\parcm\ to 12\parcm. More recently, \planck\ provide the full sky maps in total nine frequency bands ranging from 23 \GHz\ to 857 \GHz\ with beam size ranges from 32\parcm\ to 5\parcm. The last two channels of the \planck\ (545 and 857 \GHz) are not polarization-sensitive and mainly designed for intensity observation. All these multi-frequency maps are the mixture of cosmological, Galactic and extra-galactic components (e.g., CMB anisotropies, thermal dust, synchrotron, spin dust/Anomalous Microwave Emission (AME), faint/strong radio and infrared sources, thermal/kinetic Sunyaev-Zeldovich (tSZ/kSZ) effects etc.). However, for polarization, the spectrum is less complex. The high-frequency ends of the spectrum are dominated by thermal emission from Galactic dust\citep{planck-XXI:2015}. Low-frequency bands are synchrotron dominated. In addition to these, hints of polarized AME has been found \citep{Leitch:1997,Finkbeiner:2004}. However, it seems that this component plays an important role at 10-60 \GHz\ \citep{de_Oliveira-Costa:2004}, and it has a low polarization degree (1-2\%, \cite{Genova-Santos:2017}).

Separating the astrophysical sources is a crucial step in the scientific exploitation of such rich data. Over the past few years, the study of the Galactic thermal dust and synchrotron has been tied up with observational cosmology \citep{Hazumi:2019, SO:2020, CMB-S4:2016, PICO:2019} that is searching for primordial B-mode polarization in CMB, a proof of epoch of inflation \citep{Guth:1981}. The reason for this entanglement is that the expected B-mode signal in CMB imprinted from the primordial Gravitational waves during inflation is highly obscured by polarized Galactic emissions of thermal dust and synchrotron \citep{planck-I:2020}. The level of contamination depends on the energy scale of inflation \citep{Knox:2002}. Therefore, the separated foreground maps will help in building accurate modelling of thermal dust and synchrotron polarization models \citep{T_Ghosh:2017, Adak:2019, Guillet:2018, Regaldo:2020, Clark:2019, Fauvet:2011} in this regard. Furthermore, the component maps will help in detailed understanding of thermal dust and synchrotron emission, Galactic magnetic field, Galactic astrophysics etc. 
Several component separation methods have been developed over the past decades to clean the CMB signal from foregrounds, systematic effects, extra-galactic emissions. For intensity data the widely used techniques in \wmap\ and \planck\ mission are ILC \citep{Tegmark:1997}, \smica\ \citep{Delabrouille:2003}, \COMMANDER\ \citep{Eriksen:2008}, \NILC\ \citep{Basak:2011}, \SEVEM\ \citep{Fernandez-Cobos:2012}, \SILC\ \citep{SILC:2016}, \LGMCA\ \citep{LGMCA:2014} and many more to clean CMB temperature from others contamination. Out of these methods, \COMMANDER\ is a Bayesian fitting technique that can provide all astrophysical foreground maps along with the cleaned CMB map. A generalized version of Needlet ILC called \GNILC\ \citep{planck-XLVIII:2016} estimate the thermal dust maps disentangling from other Galactic foregrounds and Cosmic Infrared Background emission. Not all of these methods mentioned above provide foreground polarization maps. An updated version of \smica, \COMMANDER\ and \GNILC\ can only provide polarized thermal dust and synchrotron maps. 

Our interest lies in applying the ILC method in separating thermal dust and synchrotron polarization templates using multi-frequency data. The standard ILC method is extensively used to recover the CMB temperature maps by a weighted sum of multi-frequency data \citep{Tegmark:1997, Basak:2011, Eriksen:2004}. This paper presents another way of application of ILC aiming to estimate the foreground signals for which the electromagnetic spectrum is known. The simplicity of ILC is that it does not assumes anything about the model of the components. ILC estimates the weights by minimizing the variance of the resulting map. The minimization is generally done either in pixel space \citep{Tegmark:1997} or in harmonic space \citep{Kim:2009}. This method is only applicable to the spin-0 fields where quantities are not projected in local frames. However, in the case of polarization, we need to deal with the components having polarization vectors projected in the local frame. Stokes \Qm and \Um are not projected in a global reference frame like temperature. The mean and variance for individual spinorial components, therefore, are not defined. Therefore, a natural extension of ILC in the individual \Qm, \Um field is not possible. The straightforward way to apply a similar version of the ILC method for polarization data is to work on E- and B- mode maps \citep{Basak:2013}. However, only partial sky polarization data are commonly available in a real scenario, and decomposing them to E- and B- maps is not a trivial task. \cite{PILC:2016} develop an algorithm generalizing the standard ILC method, which applies to \Qm $\pm$ i\Um, called polarization ILC (PILC). Although \Qm $\pm$ i\Um transforms like spin-2 variable \citep{Hu_and_White:1997}, since PILC approach is based on minimization of covariant quantity, it preserves the coherence of the spinorial description. The performance of the standard ILC has limitations. It assumes all components are specially uncorrelated, whereas the Galactic foregrounds are not. For example, polarized thermal dust and synchrotron are found to be correlated \citep{Choi_and_Page:2015}. However, adding multiple constraints to reduce the contamination from other astrophysical components can significantly improve the standard ILC's performance. This method is called constrained ILC (cILC, \cite{Remazeilles:2011}). \cite{Remazeilles:2011} use this method for simultaneous estimation of CMB and thermal Sunyaev–Zeldovich emission. \cite{Hurier_2013} present this method in a more general form.   

In this paper, we develop an algorithm combining the extended version of the PILC method for heavily constraint equations (similar to cILC) with the recently developed moment expansion method of the foregrounds modelling in \cite{Chluba:2017}. Moment expansion is a powerful approach proposed by \cite{Chluba:2017} to describe the unknown complexity of the foregrounds due to variations of the spectral properties along the line-of-sight (LOS) inside the beam and across the sky. In short, moment expansion is a perturbative approach of foreground modelling under some assumption of spectral energy distribution (SED) of the components. Therefore, our method is a semi-blind component separation algorithm that performs in interface of the blind and parametric component separation methods. In the current paper, we aim to demonstrate the performance of this algorithm in  estimation of thermal dust and synchrotron \Qm, \Um maps at 353 \GHz\ and 30 \GHz\ respectively. We use three sets of \wmap\, and \planck\ simulated maps with varying foreground complexity. The purpose of using different set of simulations is to check the robustness of the algorithm independent of complexity of the foreground model.  A similar method has been applied in \cite{Remazeilles:2020} for CMB B-mode recovery , mapping relativistic tSZ effect in \cite{Remazeilles_chluba:2020} and recovery of spectral distortion signal in \cite{Rotti:2020}. Besides, we anticipate that a similar method can also be applicable in global 21 cm signal recovery.   

The paper is organized as follows. In Sect.~\ref{sec:data}, we describe the simulated data sets and binary mask used in the paper. Section.~\ref{sec:method} summarizes the methods applied in the analysis. In Sect.~\ref{sec:srategy}, we explain the strategy of implementing the method discussed in Sect.~\ref{sec:method}. In Sect.~\ref{sec:opt_sim}, we discuss the main results. Finally in Section.~\ref{sec:conclusion}, we conclude the results.

\section{Data used}
\label{sec:data}
In this section, we describe the Galactic mask and simulated data used in this paper.

\subsection{Global Mask used}
\label{sec:mask}
\begin{figure}
    \centering
    \includegraphics[width=8.4cm]{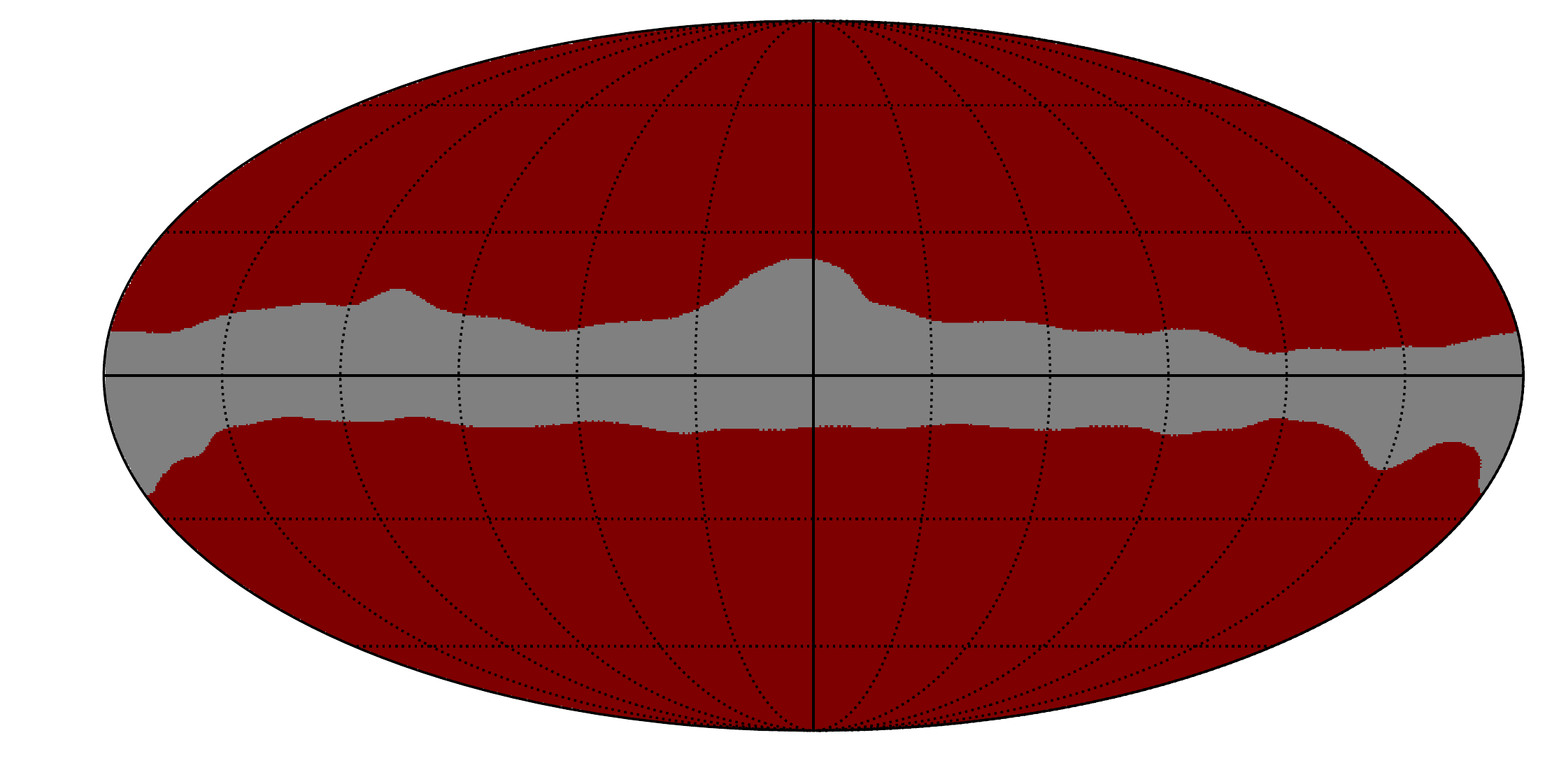}
    \caption{The \GAL\ mask that comprises 78\% of the sky. The masked region is shown in grey color and the sky used for analysis in this paper is shown in red.}
    \label{fig:mask}
\end{figure}

Due to the anisotropic nature of the foreground contributions, the application of the ILC method over whole sky data is not the most efficient way. Therefore, we use the intermediate to high Galactic region in the analysis. We use 78\% Galactic mask publicly available in Planck Legacy Archive\footnote{\url{pla.esac.esa.int/pla}}. The mask is provided in  \healpix\footnote{\url{https://healpix.jpl.nasa.gov/}} \citep{Gorski:2005} grid at \Nside\ = 2048. We downgrade the mask at \Nside\ = 256. In Figure.~\ref{fig:mask}, we present the Galactic mask at \Nside\ = 256. Hereafter, we call this mask \GAL. 
 
\subsection{Simulated data}
\label{sec:sim}
We use PySM\footnote{\url{https://github.com/bthorne93/PySM_public}} \citep{pysm:2017} for simulating Stokes IQU maps. We use $K, Ka, Q, V, W$ \wmap\ bands and \planck\ all Low-frequency instrument (LFI, \cite{Mennella:2011}) bands and High-frequency instrument (HFI, \cite{planck-IV:2011}) polarization-sensitive bands in simulations. The maps are smoothed at a common resolution of FWHM = 1$^{\deg}$ and projected at \healpix\ grid at \Nside\ = 256. We consider CMB, thermal dust, synchrotron, AME and instrument noise in all simulations. We express the final maps in Rayleigh-Jeans (RJ) unit. For CMB, we use realization of fully lensed maps at tensor-to-scalar ratio $r$ = 0.0. The values of the cosmological parameters are motivated from recent \planck\ determined values reported in \cite{planck-VI:2018}. The \wmap\ noise RMS ($\sigma_0$) for polarization are 1435, 1472, 2197, 3141, 6560 $\muK$ at $K, Ka, Q, V, W$ bands respectively. We compute noise RMS at each pixels following $\sigma_{w} (p) = \sigma_0/\sqrt{N_{obs}} (p)$, where $N_{obs} (p)$ is the \wmap\ scanning pattern at \Nside\ = 512. Finally, we simulate white noise maps from $\sigma_{w} (p)$ maps at \Nside\ = 512, smooth them at FWHM = 1$^{\deg}$ and downgraded at \Nside\ = 256. We use FFP10 noise maps \citep{planck-x:2016} for \planck\ frequencies that are available in PLA. We use $\ensuremath{\tt a2}$ model (AME is denoted by $\ensuremath{\tt a}$) for simulating AME, where 2\% global polarization is introduced as described in \citep{pysm:2017}. We finally prepare following three sets of simulations with different thermal dust and synchrotron model in PySM which we describe below. 
\begin{itemize}
\item 
 SET1: We use PySM $\ensuremath{\tt d1s1}$ model, where thermal dust and synchrotron is denoted by $\ensuremath{\tt d}$ and $\ensuremath{\tt s}$ respectively and corresponding base models are described in \cite{pysm:2017}. In $\ensuremath{\tt d1s1}$ model, PySM follow modified blackbody (MBB) for thermal dust and power-law for synchrotron. In $\ensuremath{\tt d1s1}$ model, PySM use \COMMANDER\ recovered thermal dust template at 353 \GHz\ \citep{planck-x:2016} and \wmap\ 23 GHz map \citep{Bennett:2013} as synchrotron template for polarization. Thermal dust temperature and spectral index map used here is derived using \COMMANDER. Synchrotron spectral index map is taken from \cite{Miville-Desch:2008}. \\

\item
 SET2: We use PySM $\ensuremath{\tt d4s3}$ model. This model uses a two-component thermal dust model with the templates derived in \citep{Meisner:2014}. $\ensuremath{\tt s3}$ follows a curved power-law model with a baseline curvature value of -0.052 at 23 \GHz.  \\

\item
SET2: We use PySM $\ensuremath{\tt d7s1}$ model, where thermal dust model is replaced by dust grain characterization based model described in \cite{Brandon:2017}. 
\end{itemize}
\section{Methods}
\label{sec:method}
\subsection{Moment expansion of foreground emissions}
Foreground emissions are thought to be a superposition of the emission from individual emitting blocks that can be characterized by varying SEDs. Therefore, when we observe the sky within some beam; the line-of-sight and spatial average over SEDs are inevitable. These effects alter the spectral properties of the observed emissions. For example, although spectral properties of the synchrotron emission can be described as a power-law model for individual blocks, after averaging inside the beam, it remains no longer the power-law \citep{Remazeilles:2020}. This effect results in frequency-frequency decorrelation. Aside from the above two averaging effects, downgrading the maps at lower angular resolution also gives rise to the spectral averaging effect. 

\cite{Chluba:2017} propose moment expansion method, one unique approach of foreground modelling to take into account all of these averaging effects. In this section, we briefly describe the moment expansion method of \cite{Chluba:2017} and especially apply it to thermal dust and synchrotron SEDs. 

The Galactic foregrounds can be considered as a collection of emissions of amplitude $\delta I_{\nu}(p, s)$ from different emitting layers along each LOS. $p$ denotes the pixel and $s$ denotes distance of the layer along LOS. Let us assume that we know the form of spectral properties $f(\nu , \boldsymbol{\beta})$ of the components, where  $\boldsymbol{\beta} \equiv [{\beta}_1, {\beta}_2, .., {\beta}_n] (p, s)$ denotes the general form of spectral parameters of the component of interest (e.g, For thermal dust the spectral parameters are dust temperature $T_d (p, s)$  and spectral index $\beta_{d} (p, s)$). The spectral properties likely vary across the sky inside instrumental beam as well as along LOS. However, averaging along the LOS and inside the instrumental beam, both have physically the same effect, leading to a mixture of SEDs of the emitting layers. Considering that there are infinite layers along each LOS, we can statistically model the total observed emission $I_{\nu} (p)$\footnote{Here, by $I_{\nu} (p)$, we denote Stokes $I (p)$, $Q (p)$, $U (p)$ or $E (p)$, $B (p)$ at some frequency $\nu$. Hereafter, $p$ is the central pixel of the beam.} as overall observed amplitude $I_{\nu_{0}} (p)$ at some pivot frequency $\nu_{0}$ multiplied by statistical average of SEDs, along LOS and inside the beam, $f(\nu , \boldsymbol{\beta} (p))$: 

\begin{align}
    \label{eq:fg}
    I_{\nu} (p) = I_{\nu_{0}} (p) f(\nu , \boldsymbol{\beta} (p ))
\end{align}
As shown in \cite{Chluba:2017}, we can expand  $f(\nu , \boldsymbol{\beta} (p ))$ using multi-dimensional Taylor series as\footnote{We follow the convention:  ${ \partial\beta_1^{\, i} \partial\beta_2^{\, j}\cdots \partial\beta_n^{\,k} f \left(\nu, \, \overline{\boldsymbol{\beta}} (p)\right)} = {\partial^{i+j+..+k f(\nu, \boldsymbol{\beta})}\over{\partial\beta_1^{\, i} \partial\beta_2^{\, j}\cdots \partial\beta_n^{\,k}}}\Big|_{\overline{\boldsymbol{\beta}}}$},
\begin{align}
\label{eq:f_moment_expansion}
f(\nu ,\boldsymbol{\beta}(p))&=f (\nu, \overline{\boldsymbol{\beta}})(p)
+\sum_i (\beta_i (p) -\overline{\beta}_i) \,\partial_{{\beta}_i} f(\nu, \overline{\boldsymbol{\beta}})
\nonumber\\[-0.5mm]
&\!\!\!\!
+\frac{1}{2!}\sum_i \sum_j (\beta_i (p) -\overline{\beta}_i)(\beta_j (p) -\overline{\beta}_j) \,\partial_{{\beta}_i}\partial_{{\beta}_j} f (\nu , \overline{\boldsymbol{\beta}})
\nonumber\\[-0.5mm]
&\quad+ \ldots,
\end{align}
where $\overline{\boldsymbol{\beta}} \equiv [\overline{\beta}_1, \overline{\beta}_2, .., \overline{\beta}_n]$ is the pivot value of the SED vector.

The moment map of order $i + j + ...+ k$ is defined in \cite{Chluba:2017} as:
\begin{flalign}
\label{eq:moment}
m_{ij...k} (p)
= I_{\nu_{0}} (p){\left(\beta_1(p)-\overline{\beta}_1\right)^{i}\left(\beta_2(p)-\overline{\beta}_2\right)^{j}\cdots\left(\beta_n(p)-\overline{\beta}_n\right)^{k}\over i!j!\cdots k!}.
\end{flalign}
The beauty of this approach is that foregrounds can be expressed in terms of spatially varying moments having respective constant spectral properties across the sky which is given by,

\begin{align}
{ \partial\beta_1^{\, i} \partial\beta_2^{\, j}\cdots \partial\beta_n^{\,k} f \left(\nu, \, \overline{\boldsymbol{\beta}}\right)}.
\end{align}

One can now consider the moment maps $m_{ij...k} (p)$ as different astrophysical components of total foreground contribution in multi-frequency data. These components can easily be incorporated in cILC framework, which has been described in Sect.~\ref{sec:cILC}.

In the present work, we consider the thermal dust and synchrotron as the main polarized foreground components. We apply the moment expansion particularly for these two components described below.

It is widely accepted that the synchrotron emission follows power-law in RJ unit,
\begin{align}
f_{\rm sync}\left(\nu, \beta_s(p)\right) = \left({\nu \over \nu_s}\right)^{\beta_s(p)}, 
\end{align}
where $\beta_s(p)$ is the synchrotron spectral index map.
The thermal dust follows the MBB spectrum,
\begin{align}
f_{\rm dust}\left(\nu, \beta_d(p), T_d(p)\right) = \left({\nu \over \nu_d}\right)^{\beta_d(p)+1} {\exp\left({h\nu_d\over k_BT_d(p)}\right)-1\over \exp\left({h\nu\over k_BT_d(p)}\right)-1},
\end{align}
in RJ unit, where $\beta_d(p)$ and $T_{d} (p)$ denote dust spectral index and temperature map respectively.

Implementation of the moment expansion for synchrotron spectral parameter up to second-order yields,
\begin{align}
\label{eq:sync_moments}
I_{\rm sync, \nu}(p) &= I_{\nu_s}(p) \left( \frac{\nu}{\nu_s}\right)^{\,\overline{\beta}_s\,\left(1+\frac{\Delta \beta_s(p) }{\overline{\beta}_s}\right)}\cr
&=I_{\nu_s}(p) \bigg[ f_{\rm sync} \left(\nu,\overline{\beta}_s\right)\cr
                &+ \,\Delta \beta_s(p)\,\partial_{{\beta}_s} f_{\rm sync} \left(\nu,\overline{\beta}_s\right)
                \\ \nonumber
                &+{1\over 2} \,\,\Delta \beta^2_s(p)\,\partial^2_{{\beta}_s} f_{\rm sync} \left(\nu,\overline{\beta}_s\right)+\cdots \bigg],
 \end{align}
where $\Delta \beta_s(p)=\beta_s(p) - \overline{\beta}_s$, and
\begin{flalign}
\label{eq:sync_moments1}
f_{\rm sync} \left(\nu,\overline{\beta}_s\right) &= \left({\nu\over\nu_s}\right)^{\overline{\beta}_s},\nonumber \\
\partial_{{\beta}_s} f_{\rm sync} \left(\nu,\overline{\beta}_s\right) &= \ln\left({\nu\over\nu_s}\right)f_{\rm sync} \left(\nu,\overline{\beta}_s\right),\\ 
\partial^2_{{\beta}_s} f_{\rm sync} \left(\nu,\overline{\beta}_s\right) &= \left[\ln\left({\nu\over\nu_s}\right)\right]^2f_{\rm sync} \left(\nu,\overline{\beta}_s\right). \nonumber
\end{flalign}
Similarly, for thermal dust, the moment expansion yields,
\begin{align}
\label{eq:dust_moments}
I_{\rm dust,\, \nu}(p)=\,&I_{\nu_d}(p) \bigg[ f_{\rm dust} \left(\nu,\overline{\beta}_d\right)\cr
                &+\,\Delta \beta_d(p)\;\partial_{{\beta}_d} f_{\rm dust} \left(\nu,\overline{\beta}_d, \overline{T}_{\!d}\right)\cr
                &+\,\Delta T_d(p)\;\partial_{{T}_d} f_{\rm dust} \left(\nu,\overline{\beta}_d, \overline{T}_{\!d}\right)\cr
                &+{1\over 2}\,\,\Delta \beta^2_d(p)\;
                \partial^2_{{\beta}_d} f_{\rm dust} \left(\nu,\overline{\beta}_d, \overline{T}_{\!d}\right)\cr
                &+\,
                \Delta \beta_d(p)
                \Delta T_d(p)\;
                \partial_{{\beta}_d}\partial_{{T}_d} f_{\rm dust} \left(\nu,\overline{\beta}_d, \overline{T}_{\!d}\right)\cr
                &+{1\over 2}\,\,
                \Delta T^2_d(p)\;
                \partial^2_{{T}_d} f_{\rm dust} \left(\nu,\overline{\beta}_d, \overline{T}_{\!d}\right)\cr
                &+\cdots \bigg],
\end{align}
where $\Delta \beta_d(p)=\beta_d(p) - \overline{\beta}_d$, $\Delta T_d(p)=T_d(p) - \overline{T}_d$, and
\begin{flalign}
\label{eq:dust_moments1}
&f_{\rm dust} \left(\nu,\overline{\beta}_d, \overline{T}_{\!d}\right) = \left({\nu \over \nu_d}\right)^{\overline{\beta}_d+1} {\exp\left({\overline{x}_d}\right)-1\over \exp\left({\overline{x}}\right)-1},\nonumber\\
&\partial_{{\beta}_d} f_{\rm dust} \left(\nu,\overline{\beta}_d, \overline{T}_{\!d}\right) = \ln\left({\nu\over\nu_d}\right)f_{\rm dust} \left(\nu,\overline{\beta}_d, \overline{T}_{\!d}\right),\nonumber\\
&\partial_{{T}_d} f_{\rm dust} \left(\nu,\overline{\beta}_d, \overline{T}_{\!d}\right) = {1\over\overline{T}_{\!d}}\left[{ \overline{x}\exp \left( {\overline{x}} \right) \over \exp \left( {\overline{x}} \right) - 1} - { \overline{x}_d\exp \left( {\overline{x}_d} \right) \over \exp \left( {\overline{x}_d} \right) - 1 }\right] f_{\rm dust}\left(\nu, \overline{\beta}_d, \overline{T}_{\!d}\right),\nonumber\\
&\partial^2_{{\beta}_d} f_{\rm dust} \left(\nu,\overline{\beta}_d, \overline{T}_{\!d}\right) = \left[\ln\left({\nu\over\nu_d}\right)\right]^2f_{\rm dust} \left(\nu,\overline{\beta}_d, \overline{T}_{\!d}\right),\\
&\partial^2_{{T}_d} f_{\rm dust} \left(\nu,\overline{\beta}_d, \overline{T}_{\!d}\right) = \left[\overline{x}\coth\left({\overline{x}\over 2}\right) - \overline{x}_d\coth\left({\overline{x}_d\over 2}\right)\right] {1\over \overline{T}_{\!d}}\partial_{{T}_d} f_{\rm dust} \left(\nu,\overline{\beta}_d, \overline{T}_{\!d}\right),\nonumber\\
&\partial_{{\beta}_d}\partial_{{T}_d} f_{\rm dust} \left(\nu,\overline{\beta}_d, \overline{T}_{\!d}\right) = \ln\left({\nu\over\nu_d}\right)\partial_{{T}_d} f_{\rm dust} \left(\nu,\overline{\beta}_d, \overline{T}_{\!d}\right) \nonumber
\end{flalign}
are the moment SEDs up to second-order moment expansion. Here, $x = {h \nu\over K_B T_d}$, $x_d = {h \nu_d\over K_B T_d}$ and $\overline{x} = {h \nu\over K_B \overline{T}_d}$.

\subsection{Basics of ILC algorithm}
\label{sec:cleaning}
This section review the different methodology of implementation of ILC based algorithm, which allows us to deal with different spinorial components. First, we review the implementation of standard ILC to the temperature field (spin-0 field) in Sect.~\ref{sec:T_ILC}. In Sect.~\ref{sec:ILC}, we describe the generalization of standard ILC in the spinorial frame. Next, we review the extension of the standard ILC method for a set of constraint equations, called cILC in Sect.~\ref{sec:cILC}. Finally, in Sect.~\ref{sec:cMILC}, we describe the application of cILC in framework of moment expansion in the context of the current paper. 

\subsubsection{Temperature implementation of standard ILC}
\label{sec:T_ILC}
The total observed temperature map $T_{\nu}$(p) at frequency $\nu$ is assumed to be a combination of all astrophysical and cosmological signals,
\begin{equation}
\label{eq:T_ilc_data}
    T_{\nu} (p) =  a_{\nu} S_c (p) + n_{\nu} (p),
\end{equation}
where $S_{c} (p)$ is the cth component having electromagnetic spectrum $a_{\nu}$. Let us assume $a_{\nu}$ is constant over the sky. $n_{\nu} (p)$ contains the rest of the components and noise in temperature data at frequency $\nu$. For convenience, lets rewire the Eq.~\ref{eq:T_ilc_data} in vector form for all $N_{obs}$ channels,
\begin{equation}
    \label{eq:T_ilc_data_v}
    \textbf{T} (p) = \boldsymbol{a}{S_c} (p) + \textbf{n} (p), 
\end{equation}
where vectors $\textbf{T} (p)$ and $\textbf{n} (p)$ contains data and noise for all frequencies.  
In standard ILC framework, the estimated component is,
\begin{equation}
    \label{eq:T_weighted_sum}
    \hat{S}_c (p) = \sum_{\nu} w_{\nu} T_{\nu}(p),
\end{equation}
that has minimum variance, i.e.,

\begin{equation}
    \label{eq:T_variance}
    \frac{\partial}{\partial \boldsymbol{w}}\boldsymbol{w^T \mathscr{C} w} = 0,
\end{equation}
where $\boldsymbol{\mathscr{C}} =  \left\langle d d^{T} \right\rangle$ is the covariance matrix of dimension $N_{obs} \times N_{obs}$ of the temperature data maps and  $\left\langle .. \right\rangle$ denotes the average is taken over all pixels inside the region of interest. $w_{\nu}$ is the ILC weight at frequency $\nu$.  

For unbiased estimation of $\hat{S}_c (p)$, we must assume the ILC weights $\boldsymbol{w^{T}}$ = $(w_{1}, w_2, ... w_{N_{obs}})$ should satisfy the constraint,
\begin{equation}
    \label{eq:T_constrain}
    \boldsymbol{w^T a} = 1.
\end{equation}

Combining Eq.~\ref{eq:T_variance} and Eq.~\ref{eq:T_constrain} with Lagrange multiplier $\lambda$, we get, 

\begin{equation}
    \label{eq:T_lagrange}
    \frac{\partial}{\partial \boldsymbol{w}} \left [\boldsymbol{w^T \mathscr{C} w} + \lambda (1 - \boldsymbol{w^T a})\right] = 0.
\end{equation}
Solving the system of equation.~\ref{eq:T_lagrange}, the ILC weights are determined as,
\begin{equation}
    \label{eq:T_opt_weight}
    \boldsymbol{w}^{T} = \boldsymbol{ a^{T} \mathscr{C}^{-1} (a \mathscr{C}^{-1} a)^{-1}}. 
\end{equation}

\subsubsection{ILC in polarization}
\label{sec:ILC}
The straightforward generalization of standard ILC for polarization is application of the method described in Sect.~\ref{sec:T_ILC} on \Em- and \Bm maps decomposed from \Qm, \Um maps. Decomposition of \Qm, \Um maps to \Em- and \Bm maps over incomplete sky is not a trivial task. Because some amount of \Em-mode leaks into B-mode maps during decomposition over incomplete sky. \cite{PILC:2016} generalize the standard ILC for $P_{\nu}^{\pm}(p) = Q_{\nu} (p) \pm iU_{\nu}(p)$ maps which transform like spin-2 field. In this section, we briefly review this technique. The $P_{\nu}^{\pm}(p)$ map at frequency $\nu$ can be considered as a sum of component maps,

\begin{equation}
\label{eq:ilc_data}
    P_{\nu}^{\pm} (p) = \sum_{c = 1}^{N_{c}} {A}_{\nu c} P_{c}^{\pm} (p) + N_{\nu}^{\pm}  (p),
\end{equation}
where $P_c^{\pm}(p) = Q_c (p) \pm iU_c (p)$ indicates the spin-2 quantities of the individual components, $Q_c (p)$, $U_c (p)$ being the Stokes \Qm, \Um maps of the components. $A_{\nu c}$ is the coefficient of the \textit{mixing matrix} $\textbf{A}$. $N_{\nu}^{\pm} = Q_n (p) \pm iU_n (p)$ indicates the spin-2 field of the instrument noise at frequency $\nu$ and $N_c$ is the number of the components present in the data.

Assuming the mixing matrix is constant across the sky or over the domain of some pixels $\mathscrsfs{D} (p)$, the Eq.~\ref{eq:ilc_data} can be rewritten in vector form for all $N_{obs}$ observed channels as,

\begin{equation}
    \label{eq:ilc_data_v}
    \textbf{P}^{\pm}  (p) = \textbf{A } P_c^{\pm}  (p) + \textbf{N}^{\pm} (p) 
\end{equation}
where $\textbf{P}^{\pm} (p)$ and $\textbf{N}^{\pm} (p)$ are respectively the vectors containing data and noise spin-2 fields for all $N_{obs}$ observed channels at pixel $p$. $\boldsymbol{P_c}^{\pm}  (p)$ vector contains the spin-2 fields of the components. Mixing matrix $\textbf{A}$ has the dimension of $N_{obs} \times N_{c}$.   

\cite{PILC:2016} originally develop the method for estimating the CMB polarization maps where the spectral property of CMB is assumed to be unity in the thermodynamic unit ($K_{CMB}$). Here, we describe the method for a general component $P_c^{\pm}  (p)$ which has a spectral property $f_c$. The ILC approach demands prior information of the spectral property $f_c$ of the component of interest $P_c^{\pm}  (p)$ and estimates that component map from the weighted sum of the total frequency maps. \cite{PILC:2016} assumes these weights are the complex numbers and hence the component of interest can be estimated as,
\begin{equation}
    \label{eq:weighted_sum_pilc}
    \hat{P}_c^{\pm}  (p) = (\boldsymbol{w}^{T} \pm i\, \boldsymbol{m}^T) \boldsymbol{P}^{\pm} (p) = \sum_{\nu} (w_{\nu} \pm i\, m_{\nu}) P_{\nu}^{\pm} (p).
\end{equation}
The weights are determined from minimum variance of $ |\hat{P}_c (p)|^2 $ in such a way that spectrum $f_c$ Of the component must satisfy the following constraint equations,
\begin{flalign}
    \label{eq:constrain_pilc}
    &\boldsymbol{w^T f_c} = 1,\nonumber\\
    &\boldsymbol{m^T f_c} = 0.
\end{flalign}
A special case of the Eq.~\ref{eq:constrain_pilc} is that where $m_{\nu}$ is zero for all the frequencies. A similar approach has been described in \cite{Kim:2009}. Here, we adopt this special case instead of a more general version of the algorithm described in Sect.2.2 of \cite{PILC:2016}. Therefore, Eq.~\ref{eq:weighted_sum_pilc} gets simplified to the standard form, 
\begin{equation}
    \label{eq:weighted_sum}
    \hat{P}_c^{\pm} (p) = \boldsymbol{w}^{T} \boldsymbol{P}^{\pm} (p) = \sum_{\nu} w_{\nu} P_{\nu}^{\pm} (p),
\end{equation}
that must has minimum variance, i.e.,

\begin{equation}
    \label{eq:variance}
    \frac{\partial}{\partial \boldsymbol{w}} \left\langle |\hat{P}_c (p)|^2 \right\rangle = \boldsymbol{w^T C w} = 0,
\end{equation}
with the constraint,
\begin{equation}
    \label{eq:constrain}
    \boldsymbol{w^T f_c} = 1,
\end{equation}

where, $\textbf{C} =  \left\langle \boldsymbol{d} (p)\boldsymbol{d}^{\dagger} (p) \right\rangle$ is the covariance matrix of dimension of $N_{obs} \times N_{obs}$ of the data maps ($\dagger$ denotes conjugate transpose of the matrix) and the $\boldsymbol{w^{T}}$ = $(w_{1}, w_2, ... w_{N_{obs}})$ are the weights to the $N_{obs}$ frequency maps. The elements of the covariance matrix is computed as,
\begin{equation}
    \label{eq:cov}
    C_{\nu \nu^{'}} = \left\langle Q_{\nu} (p) Q_{\nu^{'}} (p) +  U_{\nu} (p) U_{\nu^{'}} (p) \right\rangle
\end{equation}
Note that $ \boldsymbol{d} (p)\boldsymbol{d}^{\dagger} (p)$ is a covariant quantity and hence defined in a global reference frame.
Here, $\boldsymbol{f_c}$ is related to mixing matrix \textbf{A} through $\boldsymbol{f_c = A e_c}$, where $\boldsymbol{e_c}$ is a vector of dimension $1 \times N_c$ of which all the elements are zero except the cth element that is one, $e_c = [0,0,0,..1,..0]^{T}$

The weights can be computed by solving $N_{obs}$ linear system of the equation along with Eq.~\ref{eq:constrain} using Lagrange undetermined multiplier method. A straightforward algebra yields,
\begin{equation}
    \label{eq:lagrange_multiplies}
    \begin{pmatrix}
    2\boldsymbol{C} & -\boldsymbol{f_c}\\
\boldsymbol{f_c}^T & 0  
\end{pmatrix}  \begin{pmatrix} \boldsymbol{w}\\ \lambda \end{pmatrix}\,\, = \,\, \begin{pmatrix} \boldsymbol{0}\\ 1\end{pmatrix} ,
\end{equation}

where \textbf{0} denotes the column matrices of all elements zero, and $\lambda$ is the Lagrange multiplier. Solving the system of equation.~\ref{eq:lagrange_multiplies}, we obtain the weights,

\begin{equation}
    \label{eq:opt_weight}
    \boldsymbol{w}^{T} = \boldsymbol{ f_c^{T} C^{-1} (f_c C^{-1} f_c)^{-1}}. 
\end{equation}
Finally, the estimated component map is, 
\begin{flalign}    
\label{eq:opt_component}
    \hat{P}_c^{\pm} (p)& = \boldsymbol{(f_c C^{-1} f_c)^{-1} f_c^{T} C^{-1} P^{\pm} (p)}\\ \nonumber& = P_c^{\pm} (p) + \sum_{i = 1, i\neq c}^{N_{c}-1} w_{\nu} ({A}_{\nu c} P_{i}^{\pm} (p) + N_{\nu}^{\pm} (p))\\\nonumber & = P_c^{\pm} (p) + F_c^{\pm} (p) + N_c^{\pm} (p).
\end{flalign}
The beauty of this method is that we can directly work on \Qm, \Um space over an incomplete sky. This is useful since the Galactic masks are conventionally defined in \Qm, \Um space. It is essential to use the Galactic masks. Otherwise, ILC weights will be determined mainly by the variance of the pixels at the Galactic plane.

It is important to note that the estimated map is biased due to the non-zero projection $F_c^{\pm}(p)$ of the SEDs of other components on the SEDs of the component of interest. Besides, the solution is biased by residual leakage of instrumental noise $N_c^{\pm}(p)$ and chance correlation between components. However, one can demand that the solution can be made better by minimizing the variance and optimizing the weights having a unit response to the $f_c$ and simultaneously zero response to other components' SEDs. This method is called constrained ILC, which has been described in the next section. 
\subsubsection{Constrained ILC in general form}
\label{sec:cILC}
When the emission spectra of some of the components are known, it is possible to deproject them using additional constraint equations in the variance minimization process of ILC. \cite{Remazeilles:2011} have applied this method in simultaneous estimation of CMB and tSZ components. However, in practice, we can put constraints for any number of components $N_{rc +1}$ of known SEDs as,
\begin{align}
\label{eq:set_of_constrain}
& \boldsymbol{w^{T}f_1} = 0,\nonumber\\
&\boldsymbol{w^{T}f_2} = 0,\nonumber\\
 \vdots \nonumber \\
 &\boldsymbol{w^{T}f_c} = 1,\\
 \vdots \nonumber \\
 & \boldsymbol{w^{T}f_{N_{rc}+1}} =0.\nonumber
\end{align}
Here, our goal is to estimate the cth component eliminating the contamination of selected  $N_{rc}$ components. To express the constraint equations in more general from, we can define a matrix $\textbf{F}$ of dimension $N_{obs} \times (N_{rc} + 1)$ as,

\begin{equation}
\label{eq:F_matrix}
    \boldsymbol{F} = \begin{pmatrix}
    f_1[1] & \cdots & f_{N_{rc} +1} \\
   \vdots & \ddots & \vdots \\
    f_{1}[N_{obs}]& \cdots & f_{N_{obs}}[N_{obs}]
    \end{pmatrix}.
\end{equation}
Then the set of equations.~\ref{eq:set_of_constrain}\, now can be conveniently expressed as,
\begin{equation}
    \label{eq:set_of_constrain1}
    \boldsymbol{F^{T} w} = \boldsymbol{e},
\end{equation}
where $\boldsymbol{e} = [0, 0, ... 1,..0]^{T}$ is the column matrix with all elements zero except cth element that is one. In this case, Eq.~\ref{eq:lagrange_multiplies} can be generalized to,

\begin{equation}
    \label{eq:lagrange_multiplies_gen}
    \begin{pmatrix}
    2\boldsymbol{C} & -\boldsymbol{F}\\
\boldsymbol{F}^T & 0  
\end{pmatrix}  \begin{pmatrix} \boldsymbol{w}\\ \boldsymbol{\lambda} \end{pmatrix}\,\, = \,\, \begin{pmatrix} 0\\ \boldsymbol{e}\end{pmatrix} ,
\end{equation}
where $\boldsymbol{\lambda} = (\lambda_1, \lambda_2, ..., \lambda_{N_{rc} + 1})^{T}$ is the vector containing $N_{rc} + 1$ Lagrange multipliers. Simple algebraic solution of system of equation.~\ref{eq:lagrange_multiplies_gen} gives the optimized weights as,
\begin{equation}
    \label{eq:cmilc_weights}
    \boldsymbol{w}^T = \boldsymbol{e^{T}} (F^{T}C^{-1}F)^{-1} F^{T} C^{-1}.
\end{equation}
The estimated component can be expressed as,
\begin{equation}
    \label{eq:opt_component_gen}
    \hat{P}_c^{\pm} (p) = \{ \boldsymbol{e^{T}} (F^{T}C^{-1}F)^{-1} F^{T} C^{-1}\} \boldsymbol{P}^{\pm} (p).
\end{equation}

The variance of standard ILC is less than that of cILC (See Section. 3.4 of \cite{Remazeilles:2020}). It causes a larger noise residual compared to that for standard ILC because of large constraints. However, cILC reduces the foreground residual compared to standard ILC. Therefore, we need to find the optimum number of constraints to balance the noise penalty and leakage from unconstrained components to the recovered map. 

\subsubsection{Moment based constrained ILC for estimation of dust and synchrotron maps}
\label{sec:cMILC}
We want to highlight that the zeroth-order moment maps in  Eq.~\ref{eq:sync_moments}, and Eq.~\ref{eq:dust_moments} are, in principle, the synchrotron and thermal dust templates at respective pivot frequencies. Here, we aim to estimate thermal dust and synchrotron templates at pivot frequencies of 353 \GHz\ and 30 \GHz\ respectively. For that, we make use of the cILC method for a set of constraints applied on the moment SEDs of different order in Eq.~\ref{eq:sync_moments}, and Eq.~\ref{eq:dust_moments}. In short, we are aiming to estimate the zeroth-order moment maps of thermal dust and synchrotron using the cILC framework projecting out other higher-order moments applying the orthogonality condition to higher-order moment SEDs w.r.to the SED of the zeroth-order moments of the respective components. Hereafter, we refer this method to be cMILC algorithm.

For estimating thermal dust template at 353\GHz, we adopt a subset of the following constraints in cMILC algorithm:
\begin{equation}
\label{eq:cmilc_dust}
\left.\begin{aligned}
& \bw^{\rm T} \cdot  f_{dust}\left(\nu, \overline{\beta}_d, \overline{T}_{\!d}\right) = 1 \\[1.5mm]
& \bw^{\rm T} \cdot \ba = 0\\[1.5mm]
&\bw^{\rm T} \cdot  f_{sync}\left(\nu, \overline{\beta}_s\right) = 0 \\[1.5mm]
& \bw^{\rm T} \cdot  \partial_{{\beta}_s}f_{sync}\left(\nu, \overline{\beta}_s\right) = 0\\[1.5mm]
& \bw^{\rm T} \cdot  \partial_{{\beta}_d}f_{dust}\left(\nu, \overline{\beta}_d, \overline{T}_{\!d}\right) = 0\\[1.5mm]
& \bw^{\rm T} \cdot  \partial_{{T}_d}f_{dust}\left(\nu, \overline{\beta}_d, \overline{T}_{\!d}\right) = 0\\[1.5mm]
& \bw^{\rm T} \cdot  \partial^2_{{\beta}_s}f_{sync}\left(\nu, \overline{\beta}_s\right) = 0\\[1.5mm]
& \bw^{\rm T} \cdot  \partial^2_{{\beta}_d}f_{dust}\left(\nu, \overline{\beta}_d, \overline{T}_{\!d}\right) = 0\\[1.5mm]
& \bw^{\rm T} \cdot  \partial^2_{{T}_d}f_{dust}\left(\nu, \overline{\beta}_d, \overline{T}_{\!d}\right) = 0\\[1.5mm]
& \bw^{\rm T} \cdot  \partial_{{\beta}_d}\partial_{{T}_d}f_{dust}\left(\nu, \overline{\beta}_d, \overline{T}_{\!d}\right) = 0.
\end{aligned}\right\} 
\end{equation}
Similarly, for estimating synchrotron tempalate at 30 \GHz, we simply interchange the first and third constraints in Eq~\ref{eq:cmilc_dust}:

\begin{equation}
\label{eq:cmilc_sync}
\left.\begin{aligned}
& \bw^{\rm T} \cdot  f_{sync}\left(\nu, \overline{\beta}_s\right) = 1 \\[1.5mm]
& \bw^{\rm T} \cdot \ba = 0\\[1.5mm]
&\bw^{\rm T} \cdot  f_{dust}\left(\nu, \overline{\beta}_d, \overline{T}_{\!d}\right) = 0 \\[1.5mm]
& \bw^{\rm T} \cdot  \partial_{{\beta}_s}f_{sync}\left(\nu, \overline{\beta}_s\right) = 0\\[1.5mm]
& \bw^{\rm T} \cdot  \partial_{{\beta}_d}f_{dust}\left(\nu, \overline{\beta}_d, \overline{T}_{\!d}\right) = 0\\[1.5mm]
& \bw^{\rm T} \cdot  \partial_{{T}_d}f_{dust}\left(\nu, \overline{\beta}_d, \overline{T}_{\!d}\right) = 0\\[1.5mm]
& \bw^{\rm T} \cdot  \partial^2_{{\beta}_s}f_{sync}\left(\nu, \overline{\beta}_s\right) = 0\\[1.5mm]
& \bw^{\rm T} \cdot  \partial^2_{{\beta}_d}f_{dust}\left(\nu, \overline{\beta}_d, \overline{T}_{\!d}\right) = 0\\[1.5mm]
& \bw^{\rm T} \cdot  \partial^2_{{T}_d}f_{dust}\left(\nu, \overline{\beta}_d, \overline{T}_{\!d}\right) = 0\\[1.5mm]
& \bw^{\rm T} \cdot  \partial_{{\beta}_d}\partial_{{T}_d}f_{dust}\left(\nu, \overline{\beta}_d, \overline{T}_{\!d}\right) = 0.
\end{aligned}\right\} 
\end{equation}
Here $\ba$ denotes the unit conversion factor of CMB from thermodynamic unit to RJ unit, $f_{cmb} = \frac{x_c^2e^x_c}{(e^{x_c} - 1)^2}$, where $x_c = \frac{h\nu}{k_BT_{CMB}}$ ($T_{CMB}$ = 2.7255 K). 
The matrix $\boldsymbol{F}$ in Eq.~\ref{eq:F_matrix} contains the moment SEDs. For example, for thermal dust estimation, the  matrix looks like, 
\begin{equation}
    \boldsymbol{F} = \left( \boldsymbol{f_{\rm dust}}(\nu, \overline{\beta}_d, \overline{T}_{\!d}), \boldsymbol{\ba}, \boldsymbol{f_{sync}}\left(\nu, \overline{\beta}_s\right), ....., \boldsymbol{\partial_{{\beta}_d}\partial_{{T}_d}f_{dust}}(\nu, \overline{\beta}_d, \overline{T}_{\!d}) \right)\nonumber, 
\end{equation}
with $\boldsymbol{e} = [1, 0, .....,0]^{T}$. For synchrotron estimation, columns of $\boldsymbol{f_{\rm dust}}$ and $\boldsymbol{f_{\rm sync}}$ in $\boldsymbol{F}$ interchanges. However, the dimension of the $\boldsymbol{F} $ matrix varies depending on the number of the moments passed to cMILC algorithm. As discussed in Sect.~\ref{sec:cILC}, the larger number of the constraints cause extra noise penalty; projecting out all the moments up to second-order does not ensure the estimated map is the optimized solution of the cMILC algorithm. We should make a balance between mitigation of the residual leakage from unconstrained components and degradation of noise residual through the choice of an optimum number of constraints as discussed in Sect.~\ref{sec:srategy}.

\begin{table*}[hbtp]
\caption{The list of the subsets of the SEDs passed to cMILC algorithm in different iterations for estimating dust template. The condition $\textbf{w}^T.f_{\rm dust} = 1$ is applied along with orthogonal condition to rest of the SEDs in each iteration to de-project the corresponding maps. The Ids of each of the iterations are displayed in first column. }
\label{table:dust_constrains}
\begin{tabular}{llc}
\toprule
    Id &  Subsets of moment SEDs \\
\hline
\hline
 cMILC01 &  $f_{\rm dust}$ ;  $\ba$   \\
 cMILC02 &  $f_{\rm dust}$ ;  $f_{\rm sync}$    \\
 cMILC03 &  $f_{\rm dust}$ ;  $\ba$ ; $f_{\rm sync}$ \\
 cMILC04 &  $f_{\rm dust}$ ;  $\ba$ ; $f_{\rm sync}$ ; $\partial_\beta\,f_{\rm dust}$ \\
 cMILC05 &  $f_{\rm dust}$ ;  $\ba$ ; $f_{\rm sync}$ ; $\partial_\beta\,f_{\rm sync}$ \\
 cMILC06 &  $f_{\rm dust}$ ;  $\ba$ ; $f_{\rm sync}$ ; $\partial_\beta\,f_{\rm dust}$ \\
 cMILC07 &  $f_{\rm dust}$ ;  $\ba$ ; $f_{\rm sync}$ ; $\partial_\beta\,f_{\rm sync}$ ; $\partial_\beta\,f_{\rm dust}$ \\
 cMILC08&   $f_{\rm dust}$ ;  $\ba$ ; $f_{\rm sync}$ ; $\partial_\beta\,f_{\rm dust}$  ; $\partial_T\,f_{\rm dust}$\\
 cMILC09 &  $f_{\rm dust}$ ;  $\ba$ ; $f_{\rm sync}$ ; $\partial_\beta\,f_{\rm sync}$ ; $\partial_T\,f_{\rm dust}$\\
 cMILC10 &  $f_{\rm dust}$ ;  $\ba$ ; $f_{\rm sync}$ ; $\partial_\beta\,f_{\rm sync}$ ; $\partial_\beta\,f_{\rm dust}$ ; $\partial_T\,f_{\rm dust}$ \\
 cMILC11 &  $f_{\rm dust}$ ;  $\ba$ ; $f_{\rm sync}$ ; $\partial_\beta\,f_{\rm sync}$ ; $\partial_\beta\,f_{\rm dust}$ ; $\partial_T\,f_{\rm dust}$ ; $\partial^2_T\,f_{\rm dust}$ \\
 cMILC12 &  $f_{\rm dust}$ ;  $\ba$ ; $f_{\rm sync}$ ; $\partial_\beta\,f_{\rm sync}$ ; $\partial_\beta\,f_{\rm dust}$ ; $\partial_T\,f_{\rm dust}$ ; $\partial^2_\beta\,f_{\rm sync}$ \\
 cMILC13 & $f_{\rm dust}$ ;  $\ba$ ; $f_{\rm sync}$ ; $\partial_\beta\,f_{\rm sync}$ ; $\partial_\beta\,f_{\rm dust}$ ; $\partial_T\,f_{\rm dust}$ ; $\partial^2_\beta\,f_{\rm dust}$ \\
 cMILC14 &  $f_{\rm dust}$ ;  $\ba$ ; $f_{\rm sync}$ ; $\partial_\beta\,f_{\rm sync}$ ; $\partial_\beta\,f_{\rm dust}$ ; $\partial_T\,f_{\rm dust}$ ;  $\partial_\beta\partial_T\,f_{\rm dust}$\\
 cMILC15 &  $f_{\rm dust}$ ;  $\ba$ ; $f_{\rm sync}$ ; $\partial_\beta\,f_{\rm sync}$ ; $\partial_\beta\,f_{\rm dust}$ ; $\partial_T\,f_{\rm dust}$ ; $\partial^2_\beta\,f_{\rm sync}$ ; $\partial^2_T\,f_{\rm dust}$ \\
 cMILC16 &  $f_{\rm dust}$ ;  $\ba$ ; $f_{\rm sync}$ ; $\partial_\beta\,f_{\rm sync}$ ; $\partial_\beta\,f_{\rm dust}$ ; $\partial_T\,f_{\rm dust}$ ; $\partial^2_\beta\,f_{\rm sync}$ ; $\partial_\beta\partial_T\,f_{\rm dust}$ \\
 cMILC17 & $f_{\rm dust}$ ;  $\ba$ ; $f_{\rm sync}$ ; $\partial_\beta\,f_{\rm sync}$ ; $\partial_\beta\,f_{\rm dust}$ ; $\partial_T\,f_{\rm dust}$ ; $\partial^2_\beta\,f_{\rm sync}$ ; $\partial^2_\beta\,f_{\rm dust}$ \\
 cMILC18 & $f_{\rm dust}$ ;  $\ba$ ; $f_{\rm sync}$ ; $\partial_\beta\,f_{\rm sync}$ ; $\partial_\beta\,f_{\rm dust}$ ; $\partial_T\,f_{\rm dust}$ ; $\partial^2_\beta\,f_{\rm dust}$ ; $\partial^2_T\,f_{\rm dust}$ \\
 cMILC19 &  $f_{\rm dust}$ ;  $\ba$ ; $f_{\rm sync}$ ; $\partial_\beta\,f_{\rm sync}$ ; $\partial_\beta\,f_{\rm dust}$ ; $\partial_T\,f_{\rm dust}$ ; $\partial^2_\beta\,f_{\rm sync}$ ; $\partial_\beta\partial_T\,f_{\rm dust}$ \\
 cMILC20 &  $f_{\rm dust}$ ;  $\ba$ ; $f_{\rm sync}$ ; $\partial_\beta\,f_{\rm sync}$ ; $\partial_\beta\,f_{\rm dust}$ ; $\partial_T\,f_{\rm dust}$ ; $\partial^2_\beta\,f_{\rm sync}$ ; $\partial^2_T\,f_{\rm dust}$ ; $\partial_\beta\partial_T\,f_{\rm dust}$\\
 cMILC21 & $f_{\rm dust}$ ;   $\ba$ ; $f_{\rm sync}$ ; $\partial_\beta\,f_{\rm sync}$ ; $\partial_\beta\,f_{\rm dust}$ ; $\partial_T\,f_{\rm dust}$ ; $\partial^2_\beta\,f_{\rm sync}$ ; $\partial^2_\beta\,f_{\rm dust}$ ; $\partial_\beta\partial_T\,f_{\rm dust}$\\
 cMILC22 &  $f_{\rm dust}$ ;  $\ba$ ; $f_{\rm sync}$ ; $\partial_\beta\,f_{\rm sync}$ ; $\partial_\beta\,f_{\rm dust}$ ; $\partial_T\,f_{\rm dust}$ ; $\partial^2_\beta\,f_{\rm sync}$ ; $\partial^2_T\,f_{\rm dust}$ ; $\partial^2_\beta\,f_{\rm dust}$\\
 cMILC23 & $f_{\rm dust}$ ;  $\ba$ ; $f_{\rm sync}$ ; $\partial_\beta\,f_{\rm sync}$ ; $\partial_\beta\,f_{\rm dust}$ ; $\partial_T\,f_{\rm dust}$ ; $\partial^2_\beta\,f_{\rm dust}$ ; $\partial^2_T\,f_{\rm dust}$ ; $\partial_\beta\partial_T\,f_{\rm dust}$\\
 cMILC24 &  $f_{\rm dust}$ ;  $\ba$ ; $f_{\rm sync}$ ; $\partial_\beta\,f_{\rm sync}$ ; $\partial_\beta\,f_{\rm dust}$ ; $\partial_T\,f_{\rm dust}$ ; $\partial^2_\beta\,f_{\rm sync}$ ; $\partial^2_T\,f_{\rm dust}$ ; $\partial_\beta\partial_T\,f_{\rm dust}$ ; $\partial^2_\beta\,f_{\rm dust}$\\
\bottomrule
\end{tabular}
\end{table*}
\begin{table*}
\caption{The list of the subsets of the SEDs passed to cMILC algorithm in different iterations for estimating synchrotron template. The condition $\textbf{w}^T.f_{\rm sync} = 1$ is applied along with orthogonal condition to rest of the SEDs in each iteration to de-project the corresponding maps. The Ids of each of the iterations are displayed in first column.  }
\label{table:sync_constrains}
\begin{tabular}{llc}
\toprule
    Id &  Subsets of moment SEDs \\
\hline
\hline
 cMILC01 &  $f_{\rm sync}$ ;  $\ba$   \\
 cMILC02 &  $f_{\rm sync}$ ;  $f_{\rm dust}$    \\
 cMILC03 &  $f_{\rm sync}$ ;  $\ba$ ; $f_{\rm dust}$ \\
 cMILC04 &  $f_{\rm sync}$ ;  $\ba$ ; $f_{\rm dust}$ ; $\partial_\beta\,f_{\rm dust}$ \\
 cMILC05 &  $f_{\rm sync}$ ;  $\ba$ ; $f_{\rm dust}$ ; $\partial_\beta\,f_{\rm sync}$ \\
 cMILC06 &  $f_{\rm sync}$ ;  $\ba$ ; $f_{\rm dust}$ ; $\partial_\beta\,f_{\rm dust}$ \\
 cMILC07 &  $f_{\rm sync}$ ;  $\ba$ ; $f_{\rm dust}$ ; $\partial_\beta\,f_{\rm sync}$ ; $\partial_\beta\,f_{\rm dust}$ \\
 cMILC08&   $f_{\rm sync}$ ;  $\ba$ ; $f_{\rm dust}$ ; $\partial_\beta\,f_{\rm dust}$  ; $\partial_T\,f_{\rm dust}$\\
 cMILC09 &  $f_{\rm sync}$ ;  $\ba$ ; $f_{\rm sync}$ ; $f_{\rm dust}$ ; $\partial_\beta\,f_{\rm sync}$ ; $\partial_T\,f_{\rm dust}$\\
 cMILC10 &  $f_{\rm sync}$ ;  $\ba$ ; $f_{\rm dust}$ ; $\partial_\beta\,f_{\rm sync}$ ; $\partial_\beta\,f_{\rm dust}$ ; $\partial_T\,f_{\rm dust}$ \\
 cMILC11 &  $f_{\rm sync}$ ;  $\ba$ ; $f_{\rm dust}$ ; $\partial_\beta\,f_{\rm sync}$ ; $\partial_\beta\,f_{\rm dust}$ ; $\partial_T\,f_{\rm dust}$ ; $\partial^2_T\,f_{\rm dust}$ \\
 cMILC12 &  $f_{\rm sync}$ ;  $\ba$ ; $f_{\rm dust}$ ; $\partial_\beta\,f_{\rm sync}$ ; $\partial_\beta\,f_{\rm dust}$ ; $\partial_T\,f_{\rm dust}$ ; $\partial^2_\beta\,f_{\rm sync}$  \\
 cMILC13 &  $f_{\rm sync}$ ;  $\ba$ ; $f_{\rm dust}$ ; $\partial_\beta\,f_{\rm sync}$ ; $\partial_\beta\,f_{\rm dust}$ ; $\partial_T\,f_{\rm dust}$ ; $\partial^2_\beta\,f_{\rm  dust}$  \\
 cMILC14 & $f_{\rm sync}$ ;  $\ba$ ; $f_{\rm dust}$ ; $\partial_\beta\,f_{\rm sync}$ ; $\partial_\beta\,f_{\rm dust}$ ; $\partial_T\,f_{\rm dust}$ ; $\partial_\beta\partial_T\,f_{\rm dust}$ \\
 cMILC15 &  $f_{\rm sync}$ ;  $\ba$ ; $f_{\rm dust}$ ; $\partial_\beta\,f_{\rm sync}$ ; $\partial_\beta\,f_{\rm dust}$ ; $\partial_T\,f_{\rm dust}$ ; $\partial^2_\beta\,f_{\rm sync}$ ; $\partial^2_T\,f_{\rm dust}$ \\
 cMILC16 &  $f_{\rm sync}$ ;  $\ba$ ; $f_{\rm dust}$ ; $\partial_\beta\,f_{\rm sync}$ ; $\partial_\beta\,f_{\rm dust}$ ; $\partial_T\,f_{\rm dust}$ ; $\partial^2_\beta\,f_{\rm sync}$ ; $\partial_\beta\partial_T\,f_{\rm dust}$ \\
 cMILC17 &   $f_{\rm sync}$ ;  $\ba$ ; $f_{\rm dust}$ ; $\partial_\beta\,f_{\rm sync}$ ; $\partial_\beta\,f_{\rm dust}$ ; $\partial_T\,f_{\rm dust}$ ; $\partial^2_\beta\,f_{\rm sync}$ ; $\partial^2_\beta\,f_{\rm dust}$ \\
 cMILC18 &  $f_{\rm sync}$ ;  $\ba$ ; $f_{\rm dust}$ ; $\partial_\beta\,f_{\rm sync}$ ; $\partial_\beta\,f_{\rm dust}$ ; $\partial_T\,f_{\rm dust}$ ; $\partial^2_\beta\,f_{\rm dust}$ ; $\partial^2_T\,f_{\rm dust}$ \\
 cMILC19 & $f_{\rm sync}$ ;  $\ba$ ; $f_{\rm dust}$ ; $\partial_\beta\,f_{\rm sync}$ ; $\partial_\beta\,f_{\rm dust}$ ; $\partial_T\,f_{\rm dust}$ ; $\partial^2_\beta\,f_{\rm sync}$ ; $\partial_\beta\partial_T\,f_{\rm dust}$ \\
 cMILC20 &  $f_{\rm sync}$ ;  $\ba$ ; $f_{\rm dust}$ ; $\partial_\beta\,f_{\rm sync}$ ; $\partial_\beta\,f_{\rm dust}$ ; $\partial_T\,f_{\rm dust}$ ; $\partial^2_\beta\,f_{\rm sync}$ ; $\partial^2_T\,f_{\rm dust}$ ; $\partial_\beta\partial_T\,f_{\rm dust}$\\
 cMILC21 &  $f_{\rm sync}$ ;  $\ba$ ; $f_{\rm dust}$ ; $\partial_\beta\,f_{\rm sync}$ ; $\partial_\beta\,f_{\rm dust}$ ; $\partial_T\,f_{\rm dust}$ ; $\partial^2_\beta\,f_{\rm sync}$ ; $\partial^2_\beta\,f_{\rm dust}$ ; $\partial_\beta\partial_T\,f_{\rm dust}$\\
 cMILC22 &  $f_{\rm sync}$ ;  $\ba$ ; $f_{\rm dust}$ ; $\partial_\beta\,f_{\rm sync}$ ; $\partial_\beta\,f_{\rm dust}$ ; $\partial_T\,f_{\rm dust}$ ; $\partial^2_\beta\,f_{\rm sync}$ ; $\partial^2_T\,f_{\rm dust}$ ; $\partial^2_\beta\,f_{\rm dust}$\\
 cMILC23 &  $f_{\rm sync}$ ;  $\ba$ ; $f_{\rm dust}$ ; $\partial_\beta\,f_{\rm sync}$ ; $\partial_\beta\,f_{\rm dust}$ ; $\partial_T\,f_{\rm dust}$ ; $\partial^2_\beta\,f_{\rm dust}$ ; $\partial^2_T\,f_{\rm dust}$ ; $\partial_\beta\partial_T\,f_{\rm dust}$\\
 cMILC24 &   $f_{\rm sync}$ ;  $\ba$ ; $f_{\rm dust}$ ; $\partial_\beta\,f_{\rm sync}$ ; $\partial_\beta\,f_{\rm dust}$ ; $\partial_T\,f_{\rm dust}$ ; $\partial^2_\beta\,f_{\rm sync}$ ; $\partial^2_T\,f_{\rm dust}$ ; $\partial_\beta\partial_T\,f_{\rm dust}$ ; $\partial^2_\beta\,f_{\rm dust}$\\
\bottomrule
\end{tabular}
\end{table*}

\section{Implementation strategy}
\label{sec:srategy}
We apply the cMILC algorithm in pixel space over \GAL\ mask. Since cMILC is based on the cILC algorithm, we pass the multi-frequency simulated data and different subsets of moment SEDs in different iterations. The possible subsets of moment SEDs for different iteration used in this analysis are listed in Table.~\ref{table:dust_constrains} (for thermal dust estimation) and Table.~\ref{table:sync_constrains} (for synchrotron estimation). The only difference of Table~\ref{table:dust_constrains} and Table.~\ref{table:sync_constrains} is the columns of $f_{dust}$ and $f_{sync}$ have been interchanged. To construct these moment SEDs, we should choose the pivot values of the parameters involved and pivot frequencies in moment expansion. In principle, the pivot parameters should be chosen differently in different simulations in Sect.~\ref{sec:sim}. In fact, pivot parameters should also be changed when we are using higher-order moments to describe the data.
However, in the interest of speedy analysis, we use fixed values of pivot parameters throughout the study independent of the set of simulations used. 
 We adopt the pivot synchrotron spectral index, $\overline{\beta}_s$ = -3.00 \citep{Miville-Desch:2008, Krachmalnicoff:2018, Kogut:2007}. For thermal dust, we adopt the pivot dust temperature, $\overline{T}_d$ = 19.4 K \citep{planck-XLVIII:2016} and dust spectral index, $\overline{\beta}_d$ = 1.53 \citep{planck_XI:2014, planck-x:2016, planck-XI:2018}. We choose the pivot frequencies for the synchrotron and thermal dust are $\nu_s$ = 30 \GHz\ and $\nu_d$ = 353 \GHz\ respectively. 
 
After implementing the cMILC algorithm for each of the iterations listed in in Table.~\ref{table:dust_constrains} (Table.~\ref{table:sync_constrains}) with corresponding subset of moment SEDs, we apply the cMILC weights to the total frequency maps to estimate the thermal dust map at 353 \GHz\ (synchrotron map at 30 \GHz). Our simulations are absolutely calibrated (unlike \planck\ and \wmap\ data) and hence do not attach any additional frequency-dependent terms with component maps except their respective SEDs. To assess the residual leakage from noise, we apply the same weights to the input noise maps. To evaluate the residual leakage from CMB, AME and other unconstrained higher-order moments of thermal dust and synchrotron (hereafter, we refer them together by \textit{moment residual}), we apply same weights to these components as well. In summary, the algorithm returns the dust map at 353 \GHz\ and synchrotron at 30 \GHz\ along with corresponding maps of moment residual and noise residual for different iterations simply by interchanging the first and third constrains in a set of Eq.~\ref{eq:cmilc_dust}.  

\section{Results}
\label{sec:opt_sim}
In this section, we investigate the cMILC results of recovered thermal dust and synchrotron maps to demonstrate the performance of the method. In this section, we present the results for the simulation in SET1 only. The similar results for rest of the simulations are presented in Appendix.~\ref{sec:other_sim_results}. 
\begin{figure*}
\begin{multicols}{2}
    \includegraphics[width=\linewidth]{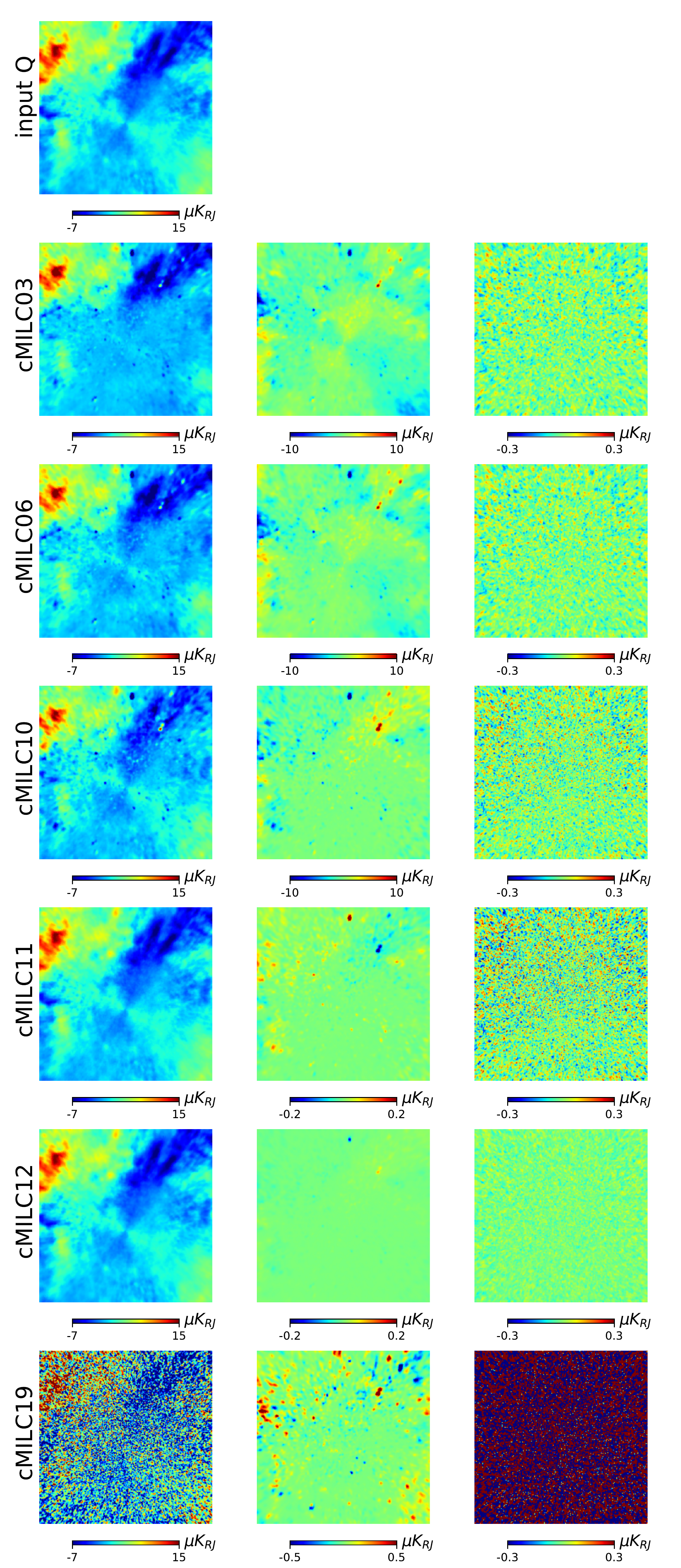} \par
    \includegraphics[width=\linewidth]{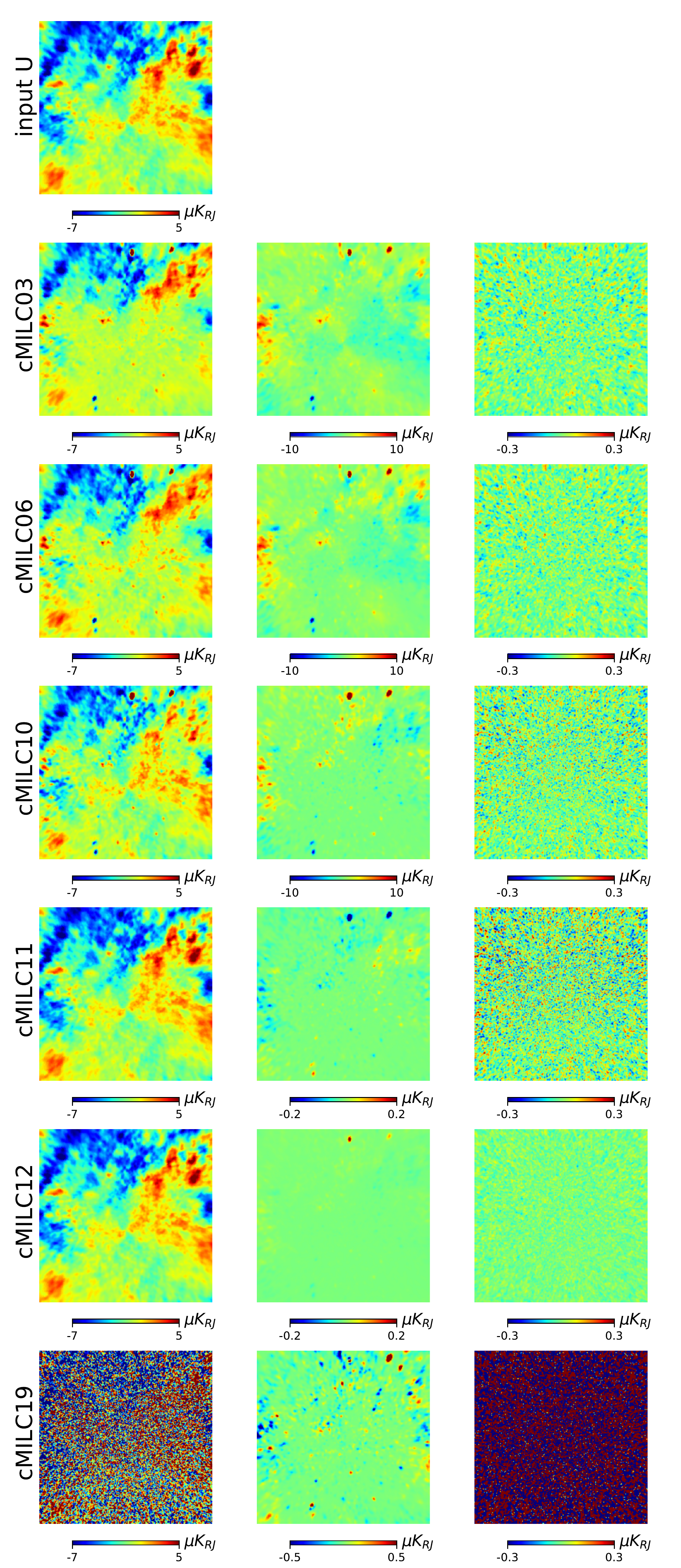} \par
    \end{multicols}
    \caption{cMILC results of estimation of thermal dust template for different iterations when deprojecting more and more moments with increasing constraints for the simulation in SET1. \textit{Left panel} shows the results of thermal dust \Qm maps, and \textit{right panel} shows the results of thermal dust \Um maps. The patches are 70$^{\deg}$ $\times$ 70$^{\deg}$ shown in gnomonic projection centered at $(l, b)$ = (90$^{\deg}$, -80$^{\deg}$). All maps are smoothed at a resolution of FWHM = 60\parcm.  The first row shows the input thermal dust map. The first, second and third columns of the subsequent rows show the recovered thermal dust maps, moment residual maps and noise residual maps respectively for some selected cMILC iterations starting from cMILC03 to cMILC19. Moment residual reduces significantly with deprojection of more and more higher-order moments up to an optimum choice of constraints till cMILC12. After that, residual increases with increasing constraints. Among all these maps, cMILC12 gives the best recovered maps.}
    \label{fig:dust_maps_sim_d1s1}
\end{figure*}

\begin{figure*}
\begin{multicols}{2}
    \includegraphics[width=\linewidth]{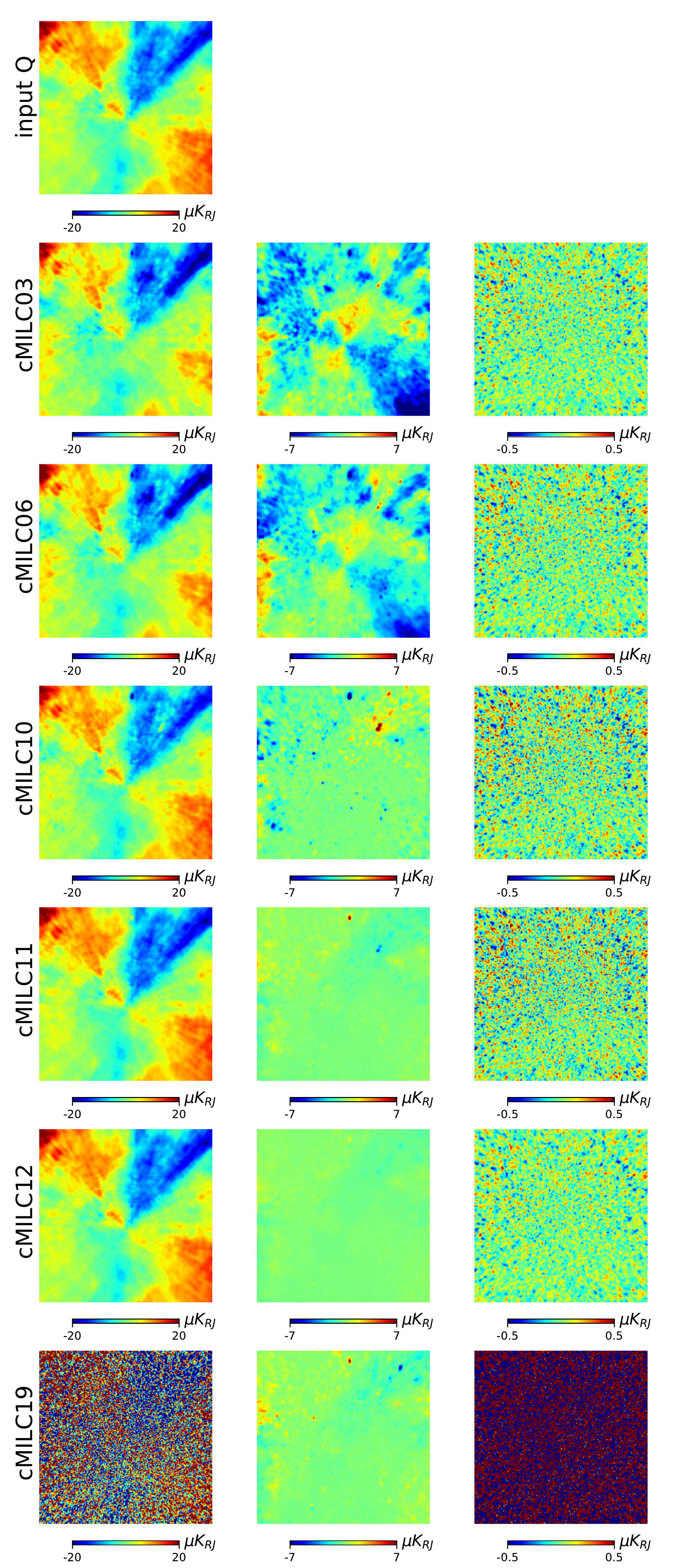} \par
    \includegraphics[width=\linewidth]{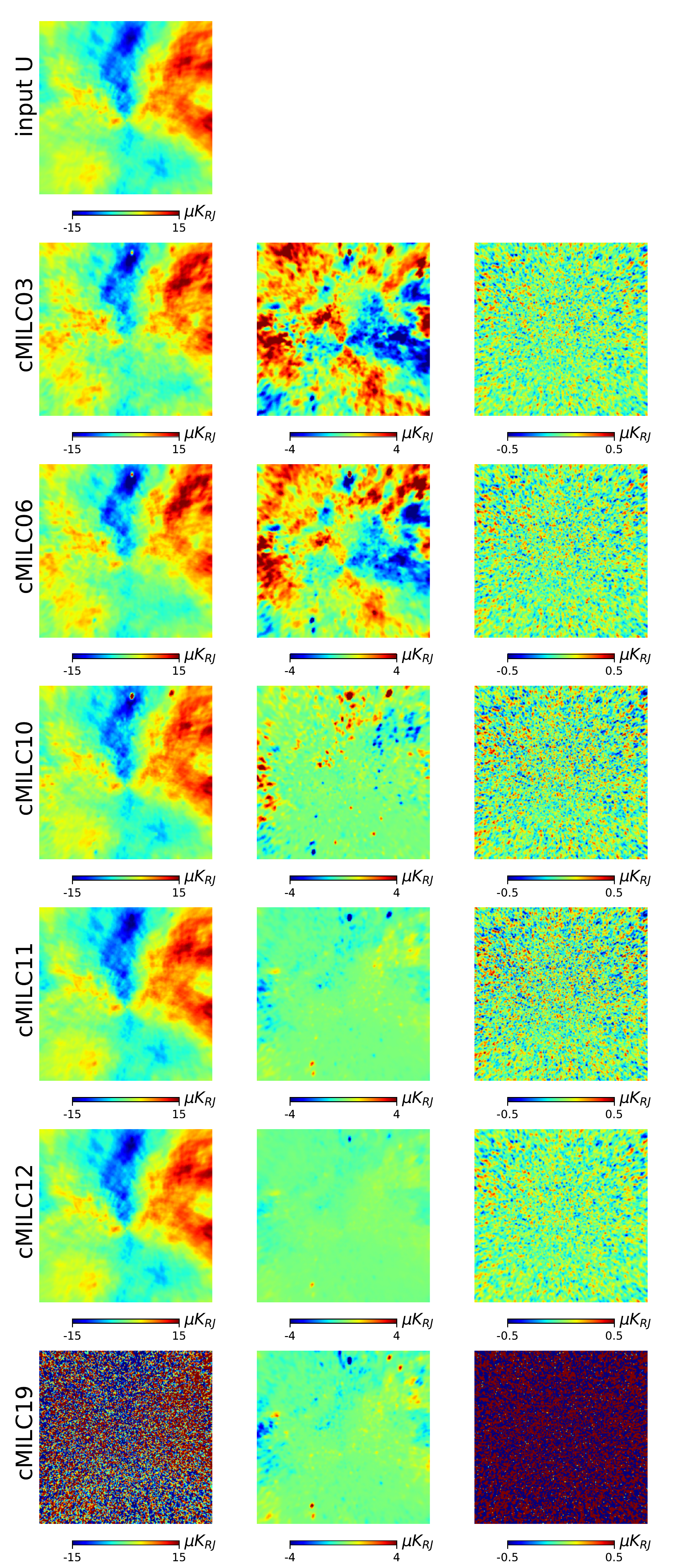} \par
    \end{multicols}
    \caption{cMILC results of estimation of synchrotron  template for different iterations when deprojecting more and more moments with increasing constraints for the simulation in SET1. \textit{Left panel} shows the results of synchrotron \Qm maps, and \textit{right panel} shows the results of synchrotron \Um maps. The patches are 70$^{\deg}$ $\times$ 70$^{\deg}$ shown in gnomonic projection centered at $(l, b)$ = (90$^{\deg}$, -80$^{\deg}$). All maps are smoothed at a resolution of FWHM = 60\parcm.  The first row shows the input synchrotron map. The first, second and third columns of the subsequent rows show the recovered synchrotron maps, moment residual maps and noise residual maps respectively for some selected cMILC iterations starting from cMILC03 to cMILC19. Moment residual reduces significantly with deprojection of higher-order moments up to an optimum choice of constraints till cMILC12. After that, residual increases with increasing constraints. Among all these maps, cMILC12 gives the best recovered maps.}
    \label{fig:sync_maps_sim_d1s1}
\end{figure*}


\begin{figure*}
    \begin{multicols}{2}
    \includegraphics[width=\linewidth]{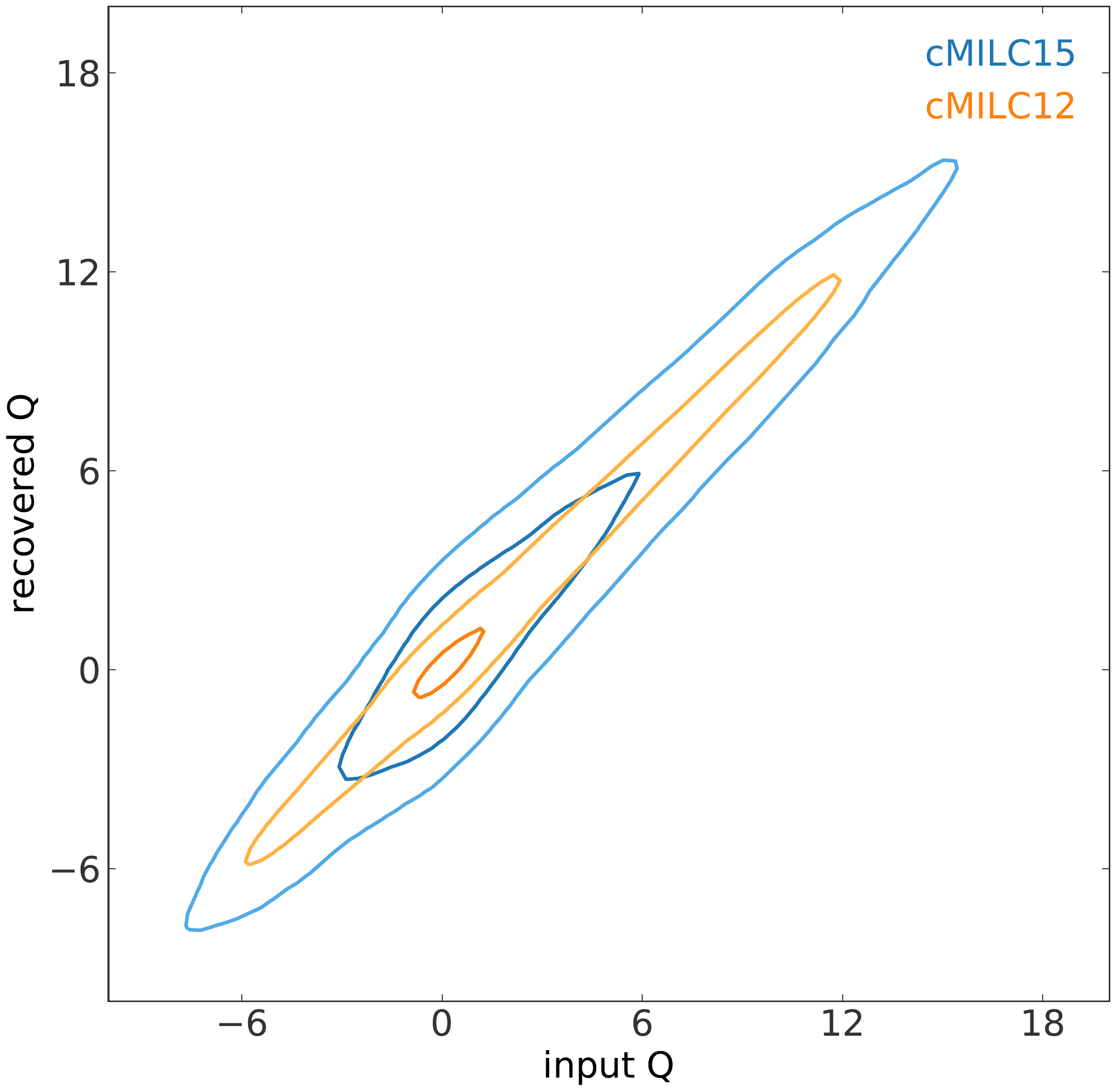}\par
    \includegraphics[width=\linewidth]{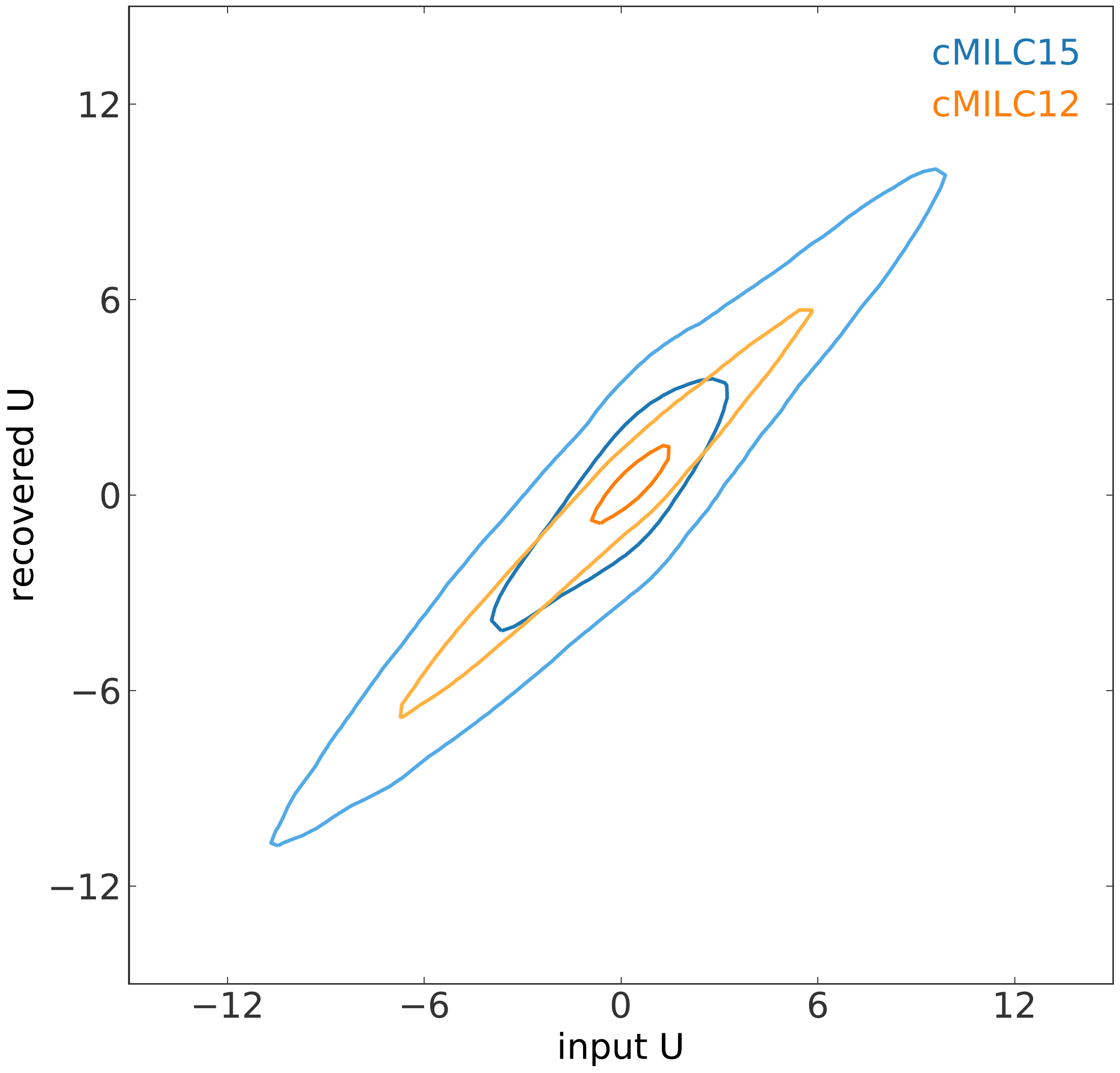}
    \end{multicols}
    
    \caption{Contour plots of 2D-histogram of input \Qm (\textit{left panel}) and \Um (\textit{right panel}) dust maps and recovered dust maps for simulation in SET1. 1$\sigma$ and 2$\sigma$ contours are shown here for cMILC12 (orange) and cMILC15 (blue) iterations. Most of the pixels are distributed inside a tiny region of distribution for output maps of cMILC12. Whereas pixels for output maps of cMILC15 are distributed inside a far bigger range of the distribution. This implies use of more than 7 constraints deteriorates the performance of algorithm for given instrument sensitivity and channels.  }
    \label{fig:dust_TT_correlation_d1s1}
\end{figure*}
\begin{figure*}
    \begin{multicols}{2}
    \includegraphics[width=\linewidth]{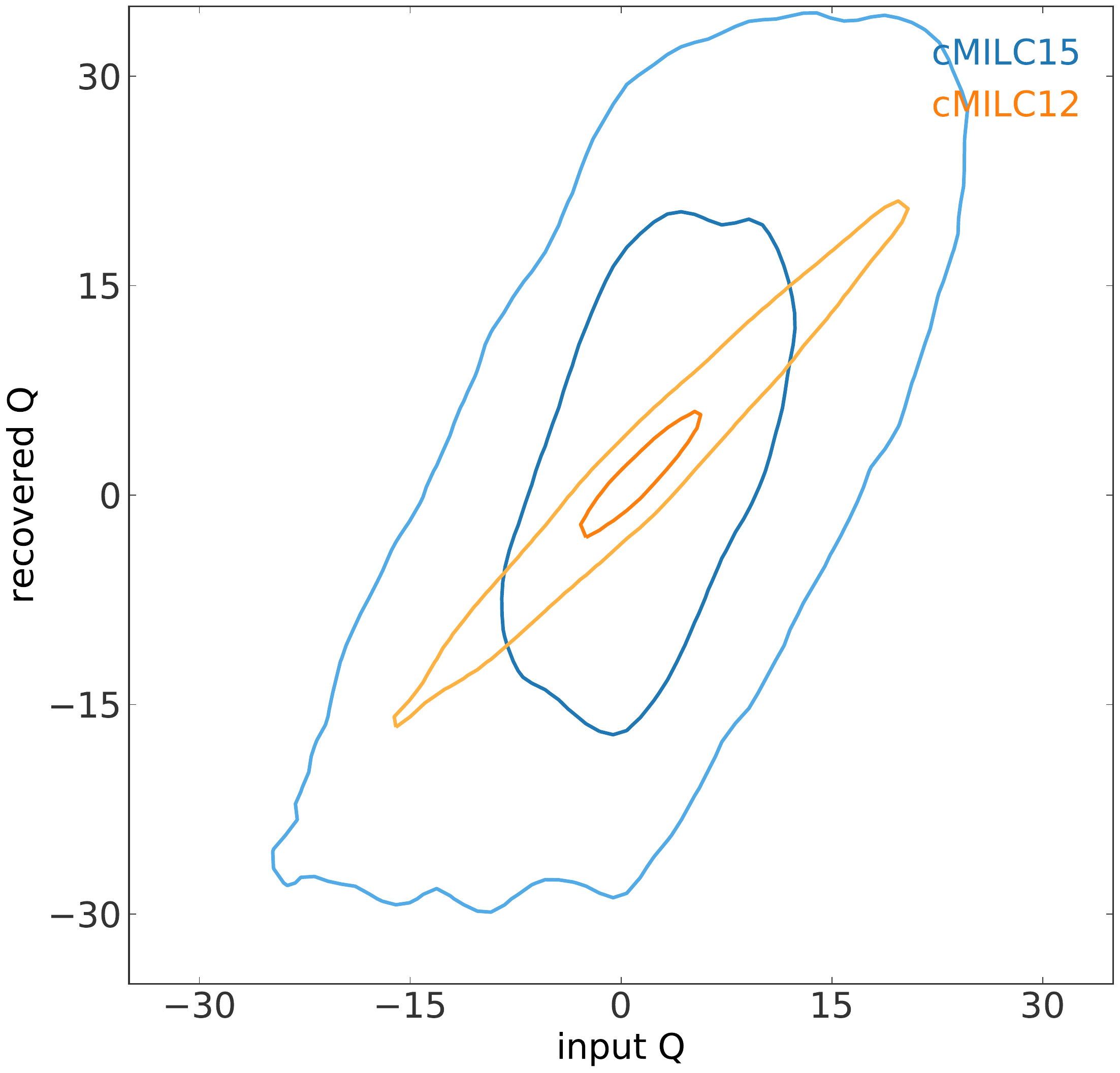}\par
    \includegraphics[width=\linewidth]{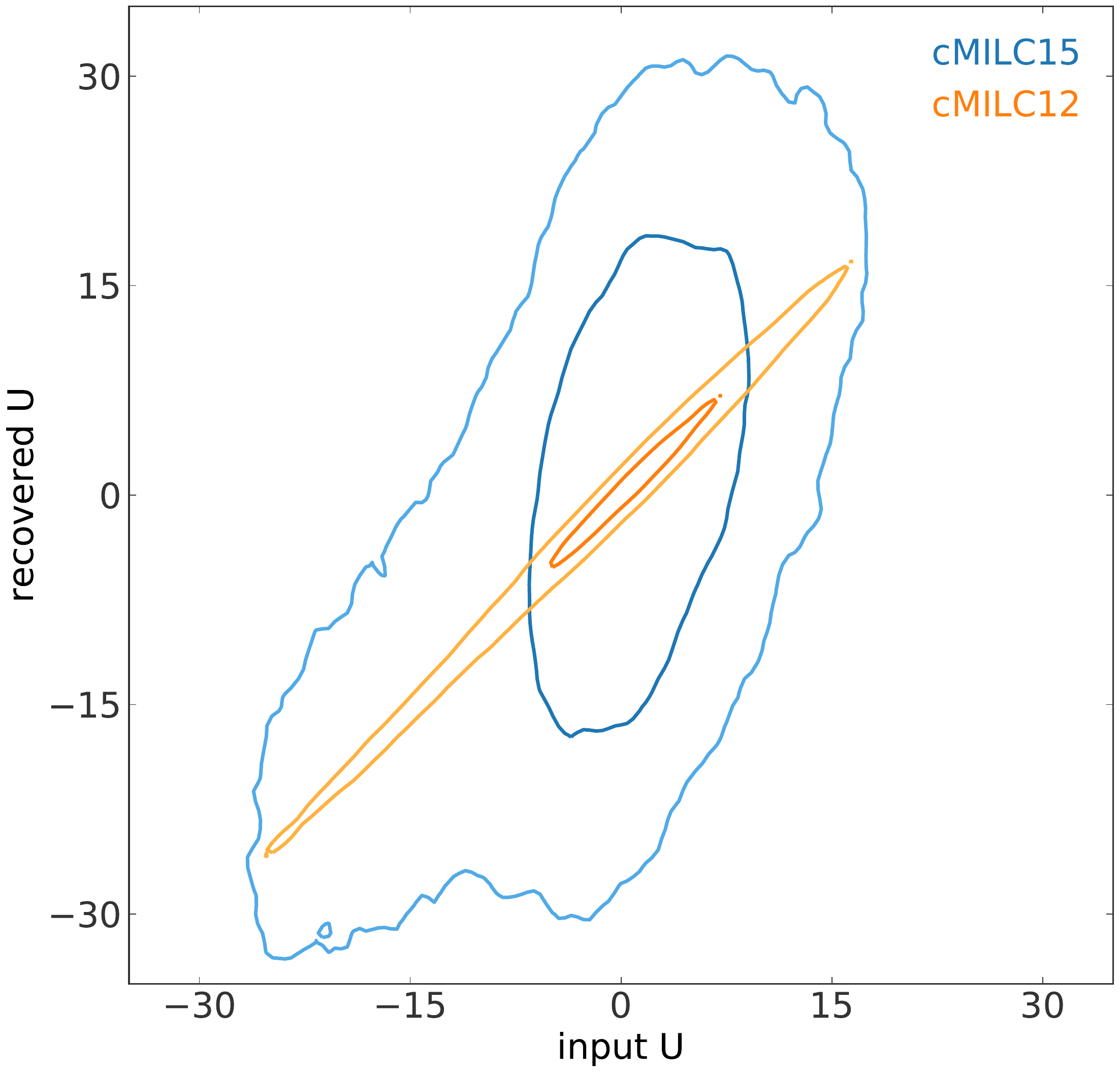}\par
    \end{multicols}

    \caption{Contour plots of 2D-histogram of input \Qm (\textit{left panel}) and \Um (\textit{right panel}) synchrotron maps and recovered synchrotron maps for simulation in SET1. 1$\sigma$ and 2$\sigma$ contours are shown here for cMILC12 (orange) and cMILC15 (blue) iterations. Most of the pixels are distributed inside a tiny region of distribution for output maps of cMILC12. Whereas pixels for output maps of cMILC15 are distributed inside a far bigger range of the distribution. This implies use of more than 7 constraints deteriorates the performance of algorithm for given instrument sensitivity and channels. }
    \label{fig:sync_TT_correlation_d1s1}
\end{figure*}

\begin{figure}
    \includegraphics[width=9cm]{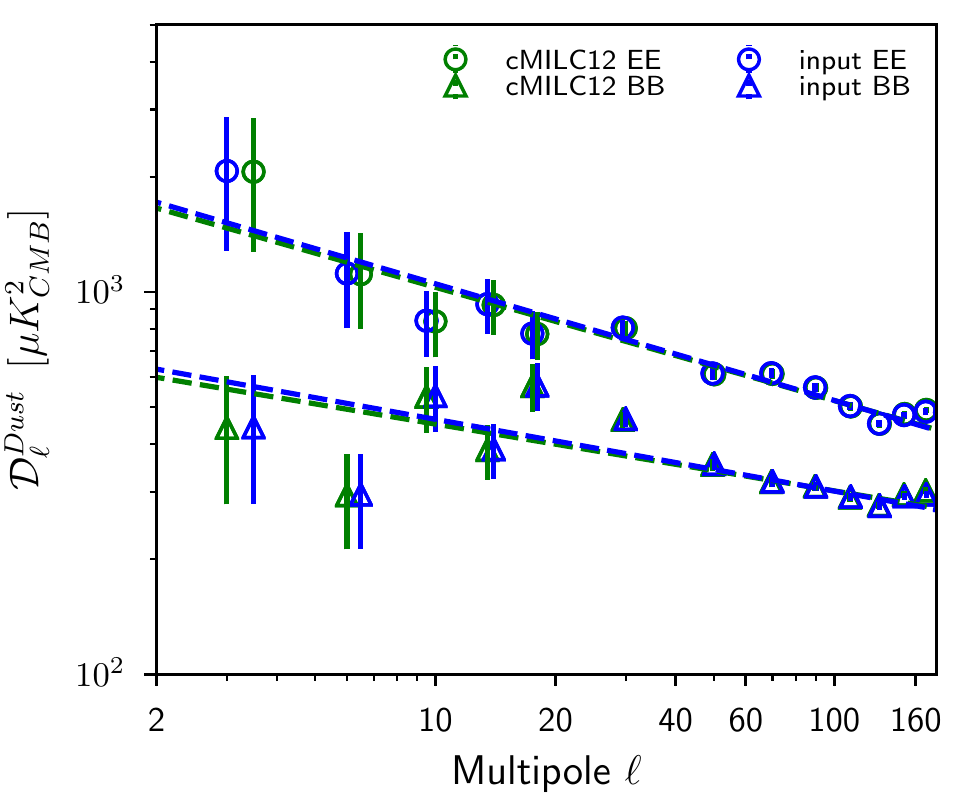}\par
    \includegraphics[width=9cm]{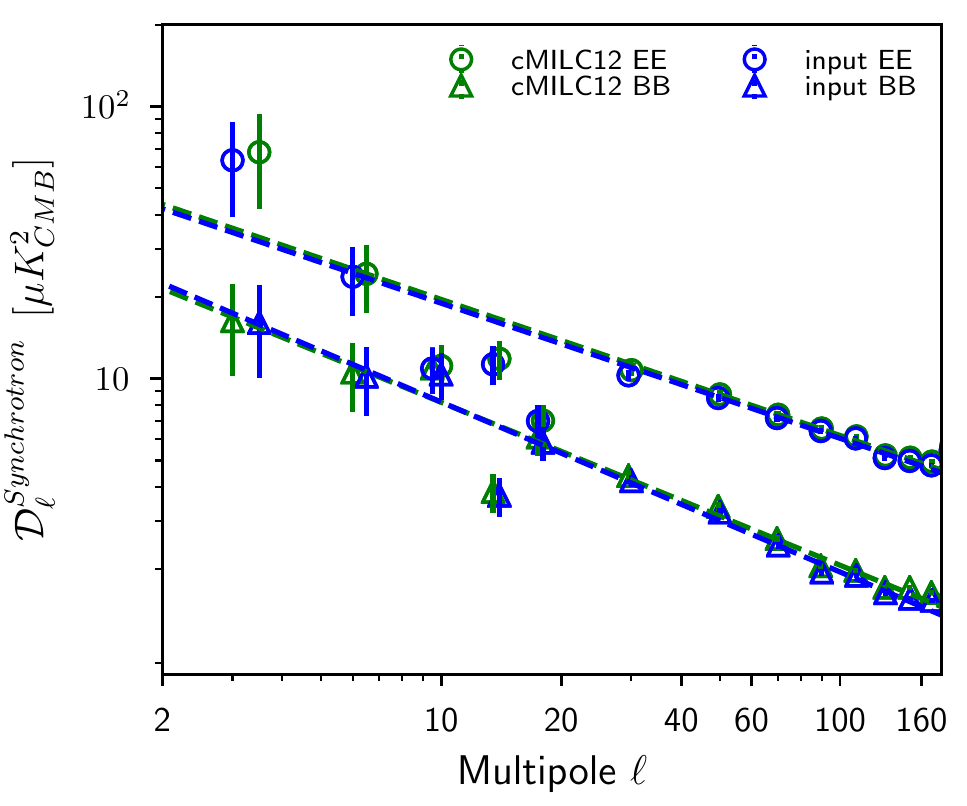} \par
    \caption{EE (circles) and BB (triangles) power spectra for thermal dust (\textit{upper panel}) and synchrotron (\textit{lower panel}) maps. Power spectra of input maps of simulation in SET1 are shown in blue, and that of recovered maps for cMILC12 iteration are shown in green. All spectra are computed over \GAL\ apodized mask using \xpol. Error bars are 1$\sigma$ uncertainties analytically computed from \xpol. The dashed lines indicate the respective best-fit power-law model power spectra. Corresponding best-fit parameters are listed in Table.~\ref{table3}.}
    \label{fig:sim_dust_sync_power_d1s1}
\end{figure}

\subsection{Inspection of recovered maps}
\label{sec:map_inspection}
We first inspect the quality of the recovered dust and synchrotron polarization maps and compare them with input maps of respective components. For illustration, we also investigate the amount of residual leakage from unconstrained components and moments as well as residual leakage of noise. 
In Figure.~\ref{fig:dust_maps_sim_d1s1}, we summarize the cMILC results of estimation of thermal dust for simulation in SET1 for some selected iterations. We display 70$^{\deg}$ $\times$ 70$^{\deg}$ patches in gnomonic projection centered at the Galactic longitude and latitude, $(l, b)$ = (90$^{\deg}$, -80$^{\deg}$). \textit{Left panel} presents the results of \Qm and \textit{right panel} presents the results of \Um. The first rows show the input thermal dust \Qm, \Um maps, the subsequent rows show the output maps at 353 \GHz\ of selected cMILC iterations that use different subset of moment SEDs. The corresponding iteration's Ids are shown on the left side of the maps. The First columns show the estimated thermal dust maps at 353 \GHz, the second columns show the moment residual maps, and the third columns show the noise residual maps. Similar results for estimation of synchrotron map at 30 \GHz\ are presented in Figure.~\ref{fig:sync_maps_sim_d1s1} over the same sky region. The cMILC03 iteration deprojects zeroth-order moments ($\ba$ ; $f_{\rm sync}$) only. Therefore, the moment residuals are reasonably high for this iteration. Deprojecting $\partial_\beta\,f_{\rm dust}$ along with zeroth-order moments (\textit{third rows}) does not reduce  the residual much to recovered maps. The moment residual reduces significantly when we deproject all zeroth- and first-order moments in cMILC10 and one of the second-order moments in cMILC11 and cMILC12. Inspecting second columns of the Figure.~\ref{fig:dust_maps_sim_d1s1} and Figure.~\ref{fig:sync_maps_sim_d1s1}, we confirm that moment residual reduces up to cMILC12 as we project out more and more moments. Inspecting the first columns, one can hardly distinguish the map-level differences in the recovered maps for cMILC03, cMILC06, cMILC10, cMILC11 and cMILC12. However, comparing the last two columns, we confirm that recovered maps for cMILC12 are the best in the sense the moment residual leakage is the least for this iteration. We also run the algorithm for simulation in absence of AME. We notice residual leakage in that case is order of magnitude less. In iterations from cMILC15 to cMILC19, we project out all the moment maps up to first order along with subsets of two second-order moments. In Figure.~\ref{fig:dust_maps_sim_d1s1} and Figure.~\ref{fig:sync_maps_sim_d1s1}, we display only the results for cMILC19 out of these four iterations where we project out two second-order moments ($\partial^2_\beta\,f_{\rm sync}$,  $\partial_\beta\partial_T\,f_{\rm dust}$) along with all zeroth- and first-order moments.
 The recovered maps in this iteration are noisy. This implies, The noise degradation for larger constrains prevents us from getting further better recovery. A similar trend in recovered maps, residual leakage from moment maps and noise have been found for other sets of simulations and shown in Appendix.~\ref{sec:other_sim_results}. Therefore, we do not inspect the rest of the iterations, which de-project more higher-order moments.

To further diagnose the recovered maps, we plot 1$\sigma$ and 2$\sigma$ contours of 2D histogram of input maps and recovered maps for cMILC12 (orange) and cMILC15 (blue) iterations in Figure.~\ref{fig:dust_TT_correlation_d1s1} (for thermal dust) and Figure.~\ref{fig:sync_TT_correlation_d1s1} (for synchrotron). We find, most of the pixels are distributed inside a very tiny region distribution for recovered maps of cMILC12 compared to that of cMILC15. Also the correlation between input and recovered maps are significanly better for cMILC12 than that of cMILC15. We find the correlation coefficients between input thermal dust maps and estimated thermal dust of cMILC12 and cMILC15 iterations are 0.78, 0.99 (for \Qm) and 0.67, 0.99 (for \Um) respectively. Similarly, the correlation coefficients for synchrotron estimation in cMILC12 and cMILC15 iterations are 0.65, 0.99 (for \Qm) and 0.61, 0.99 (for \Um) respectively. This is another proof in support of using more than seven constraints degrades the performance of cMILC algorithm for given sensitivity and frequency coverage. 

Doing all these assessments, therefore, we note that cMILC12 provides the best recovered thermal dust and synchrotron maps for joint analysis \wmap\ and \planck\ maps. However, this is not a generic solution for any mission. The performance of cMILC depends on the sensitivity and frequency coverage of the experiments.

\subsection{comparison of the power spectrum}
\label{sec:power_spectra}
In Figure.~\ref{fig:sim_dust_sync_power_d1s1}, we compare the angular power spectra of thermal dust (\textit{upper panel}) and synchrotron (\textit{lower panel}) maps as estimated for cMILC12 and input  maps. We compute $EE$ and $BB$ power spectra over \GAL\ apodized mask using \xpol\ \citep{Tristram:2005}. Results from input maps are shown in blue and that of recovered maps are shown in green. The $EE$ and $BB$ power spectra are presented in Figure.~\ref{fig:sim_dust_sync_power_d1s1} with circles and triangles respectively. The 1$\sigma$ uncertainties are analytically estimated using \xpol. We fit the power spectra with power-law model,

\begin{equation}
    \label{eq:power-law}
    \dlxx = A_{XX} (\ell/80)^{\alpha_{XX}+2},
\end{equation}
 where $A_{XX}$ is the best-fit amplitude at $\ell =80$, $\alpha_{XX}$ is the best-fit spectral index and $XX=\{EE, BB\}$. We use $\ell$ range of 30-160 of thermal dust power spectra and 2-140 of synchrotron power spectra for fitting Eq.~\ref{eq:power-law} with \mpfit\ routine following the same same machinery
as in \cite{planck-XI:2018}. The best-fit power-law model power spectra are shown in dashed lines in Figure.~\ref{fig:sim_dust_sync_power_d1s1}. The corresponding best-fit parameters are listed in Table.~\ref{table3}. 

Overall, we find an excellent agreement between power spectra of input and recovered maps both for thermal dust and synchrotron. All the parameters are comparable within 3$\sigma$ statistical uncertainty. Most importantly, we find the power ratio of $B$- and $E$- mode ($A_{BB}/A_{EE}$) measured both for input and recovered map is $\sim$0.56 for thermal dust, and $\sim$0.34 for synchrotron which are very similar to the corresponding values reported in \cite{planck-VI:2018}.    
\begin{table}
\caption{ Best-fit parameters of the power-law model fitted to the thermal dust and synchrotron power spectra of the input and recovered maps in cMILC12 iteration. 30 $\leq \ell \leq$ 160 range has been used for fitting thermal dust power spectra, and 2 $\leq \ell \leq$ 140 range has been used for fitting for synchrotron power spectra.} 
\label{table3}
   \begin{centering}
   \begin{tabular}{ p{3.2cm}  p{2.0cm}   p{2.0cm} }
   \hline
  parameters & input map & output map\\
    \hline
    \hline
    \textbf{thermal dust; $\ell$ = 30-160}& &\\
    $A_{EE}$&555.14 $\pm$  7.61 & 556.84 $\pm$  7.63 \\
    $A_{BB}$& 313.68 $\pm$ 4.35 &314.22 $\pm$  4.36\\
    $A_{BB}/A_{EE}$ & 0.57 $\pm$ 0.02& 0.56 $\pm$ 0.02\\
    $\alpha_{EE}$&-2.30 $\pm$  0.03&-2.31 $\pm$  0.03\\
    $\alpha_{BB}$&-2.17 $\pm$  0.03&-2.19 $\pm$  0.03\\
    \hline
    \textbf{Synchrotron; $\ell$ = 2-140} &&\\
    $A_{EE}$&6.91 $\pm$ 0.10 &6.74 $\pm$ 0.09 \\
    $A_{BB}$&2.35 $\pm$  0.03 &2.24 $\pm$  0.03\\
    $A_{BB}/A_{EE}$ & 0.34 $\pm$ 0.01 & 0.33 $\pm$ 0.01\\
    $\alpha_{EE}$&-2.50 $\pm$  0.03&-2.49 $\pm$  0.03\\
    $\alpha_{BB}$&-2.59 $\pm$  0.03&-2.62 $\pm$  0.03\\
    \hline
\end{tabular}
\end{centering}
\end{table}
\begin{figure*}
    \centering
    \includegraphics[width=18cm]{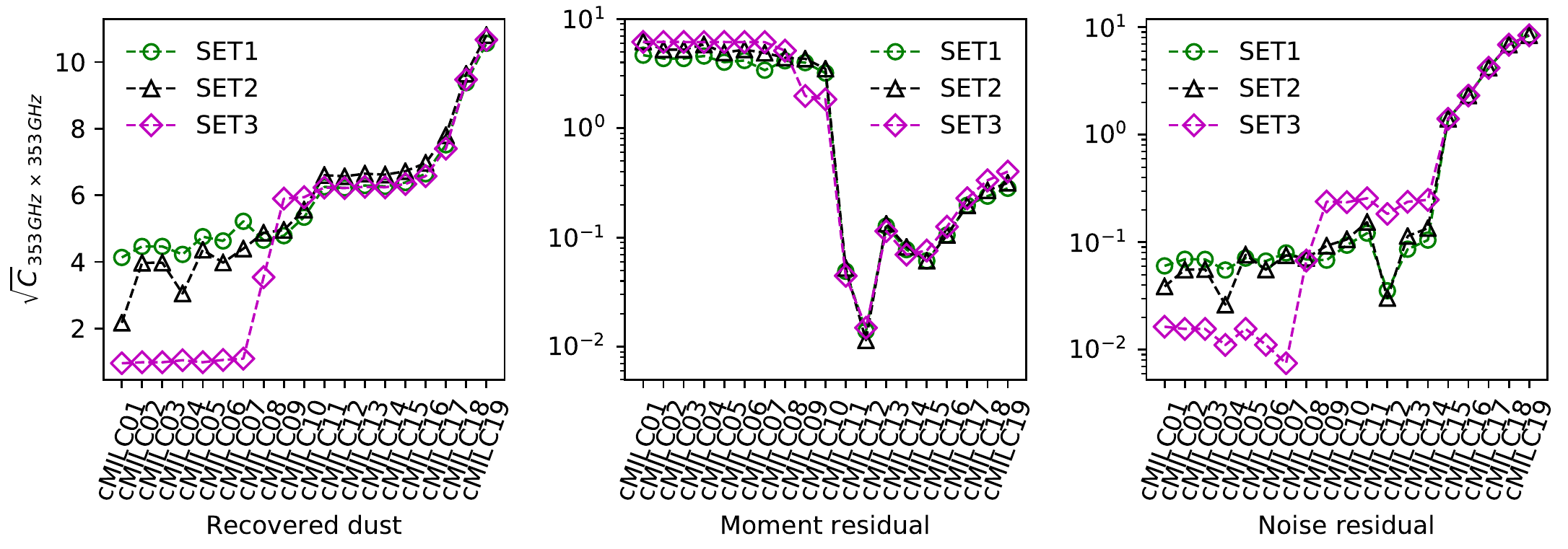}
    \caption{Evolution of the standard deviation of the output maps at 353 \GHz\ for simulation in SET1 (green), SET2 (black) and SET3 (magenta) with different cMILC iterations starting from cMILC01 to cMILC19 where we pass different subsets moment SEDs. The \textit{left panel} presents the standard deviations of the recovered thermal dust maps, \textit{middle panel} presents the standard deviations of the moment residual maps, and \textit{right panel} presents the standard deviations of the noise residual maps at 353 \GHz. }
    \label{fig:stat_res_dust}
\end{figure*}
\begin{figure*}
    \centering
    \includegraphics[width=18cm]{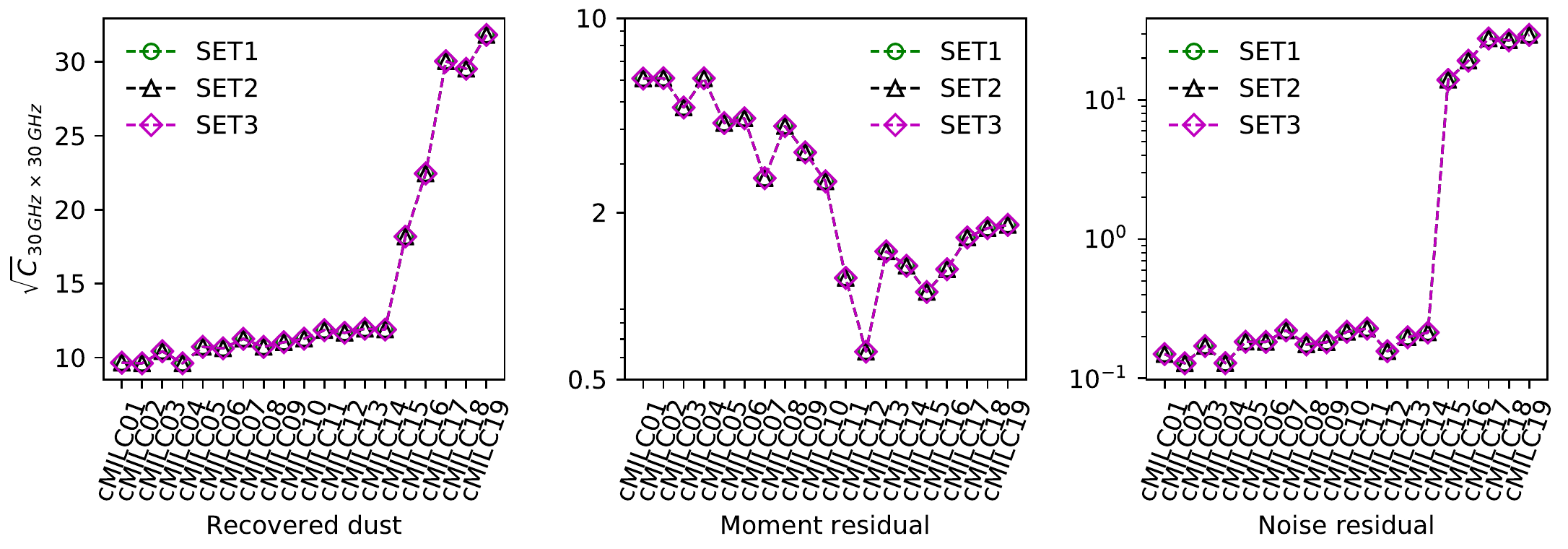}
    \caption{Evolution of the standard deviation of the output maps at 30 \GHz\ for simulation in SET1 (green), SET2 (black) and SET3 (magenta) with different cMILC iterations starting from cMILC01 to cMILC19 where we pass different subsets moment SEDs. The \textit{left panel} presents the standard deviations of the recovered synchrotron maps, \textit{middle panel} presents the standard deviations of the moment residual maps, and \textit{right panel} presents the standard deviations of the noise residual maps  at 30 \GHz.}
    \label{fig:stat_res_sync}
\end{figure*}
\subsection{Statistics of residuals from moment and noise maps}
\label{sec:stat_residuals}
Besides the map level investigation, its is also important to assess the statistical properties of the estimated maps, residual leakage from other components which are not projected out and noise residual maps. In Figure.~\ref{fig:stat_res_dust}, we present the standard deviation $\sqrt{C}_{353\, GHz \,\times \,353\, GHz}$ ($C_{\nu,\nu^{'}}$ is defined in Eq.~\ref{eq:cov} ) of the recovered thermal dust map (\textit{left panel}), residual leakage from moment maps (\textit{middle panel}) and noise residual maps (\textit{right panel}) for different cMILC iterations. Similarly, in Figure.~\ref{fig:stat_res_sync}, we present the standard deviation $\sqrt{C}_{30 \,GHz \, \times \,30\, GHz}$ of similar maps for estimation of synchrotron for different cMILC iterations. Here, we display the results for all three set of simulations for easy caparison. 

In \textit{left panel} of Figure.~\ref{fig:stat_res_dust} and Figure.~\ref{fig:stat_res_sync}, we find the standard deviations of the recovered maps are increasing with increasing number of constraints in cILC algorithm. However, for the iterations, which pass same number of constraints to cMILC algorithm but project out a different subset of moments, the standard deviations are either comparable or change. For example, standard deviations of recovered maps are approximately the same for the iterations from cMILC11 to cMILC14 which pass 7  constraints but different second-order moment SEDs along with all zeroth- and first-order moment SEDs to the cMILC algorithm. Whilst, standard deviations of recovered maps for the iterations from cMILC15 to cMILC19 changes although each of the iterations pass 8 moment SEDs to the cMILC algorithm but project out a different subset of two second-order moments along with all zeroth- and first-order moments. This implies, changes in standard deviations of the recovered maps for fixed number of constraints are subjected to the subset of moment SEDs passed to the algorithm.

Increasing standard deviation with an increasing number of constraints gives rise to a misleading expectation that projecting out more moments always come with an additional noise penalty. Third panels of Figure.~\ref{fig:stat_res_dust} and Figure.~\ref{fig:stat_res_sync} demonstrate that this is an inaccurate extrapolation. Furthermore, the reduction of the leakage from higher-order moments indefinitely with an increasing number of constraints for given sensitivity and frequency coverage is also incorrect information. On the contrary, in \textit{middle panels} of Figure.~\ref{fig:stat_res_dust} and Figure.~\ref{fig:stat_res_sync}, we find, for a given sensitivity and frequency coverage of the experiments, leakage from higher-order moments reduces up to projecting out an optimum number of moments and reaches to a minimum value. After that residual increases with projecting out more moments that is clear from \textit{middle panels} of Figure.~\ref{fig:stat_res_dust} and Figure.~\ref{fig:stat_res_sync}. 

Therefore, we would like to emphasize that the increasing number of constraints in the cMILC algorithm does not always come with noise penalty and indefinite reduction of residual from unconstrained moments in the recovered maps. It has a more complicated behaviour depending on the complexity of the foregrounds, sensitivity and frequency coverage of the mission.     

\section{Conclusion}
\label{sec:conclusion}
In the present work, we develop a new semi-blind components separation method using constrained ILC in the language of moment expansion introduced in Sect.~\ref{sec:cMILC}. We apply this algorithm to three sets of simulations with varying thermal dust and synchrotron complexity to demonstrate the performance of the algorithm.  We use \wmap\ and \planck\ instrument specification for current work. Our main objective is to estimate the zeroth-order moment maps of thermal dust and synchrotron at respective pivot frequencies 353 \GHz\ and 30 \GHz\ by projecting out the higher-order moments. The zeroth-order moment maps eventually are the individual foreground templates of respective components at respective pivot frequencies  as discussed in Sect.~\ref{sec:cMILC}. We find the best combination of the moment SEDs to project out the specific moments that optimize the trade-off between residual from unconstrained higher-order moments and noise degradation in the  templates. However, this combination is not robust and specific to the sensitivity and frequency coverage of the instruments. We show the performance of the cMILC method is optimal up to a specific number of constraints applied for given instrument sensitivity and channels. After that, the performance of algorithm deteriorates with increasing constraints since the residual bias from unconstrained moments increases.
Furthermore, we show deprojecting more and more higher-order moments does not always come with noise penalty. It depends on the combination of moment SEDs passed to the algorithm. Eventually, this aspect would be more apparent if we would work with high sensitive instrument data like PICO \citep{PICO:2019} to estimate low signal-to-noise components like B-mode signal in CMB. We do not apply constraints on AME in the present work since the moment description of this component is not available in literature. We notice that unconstrained AME introduce an extra bias that is order of magnitude high in comparison to that from unconstrained moments.       

Overall, this is a new method to estimate the foreground templates. We develop this method on spin-2 fields and can easily be extended to the spin-0 field. However, for intensity maps, lots of foreground components contribute, unlike polarization. Developing a moment description for some of the foregrounds in intensity (e.g., AME and CO line emissions) will be essential for optimal performance of cMILC algorithm. This turns into a high dimensional problem and finding the most relevant SEDs to project out using a very limited number of frequency coverage (only 12 channels is used in this work) is substantially challenging. Therefore, we do not apply this method to the intensities. However, the number of moment SEDs required for the optimal solution is directly related to the requirement of the number of frequency channels with some sensitivity. Thus algorithm can be useful for optimizing the design of the future CMB experiments.         

The algorithm we have developed works over any sky fraction. Therefore, in principle, we can jointly analyse ground-based and space-based CMB mission data using this algorithm. The most challenging parts of working with real data using this algorithm are calibration and beam uncertainties. In the present work, we assume the maps are absolutely calibrated, and Gaussian FWHM can perfectly describe beams. However, for real data, calibration coefficient uncertainties for each channel, which are a multiplicative factor for each frequency maps, introduce an uncertainty in the frequency scaling of each of the components. Therefore, the optimal combination of moment SEDs for given instrumental sensitivity and frequency coverage may converge to imperfect solution of the component maps. Beam uncertainties induce a similar bias as calibration uncertainties. This impacts strongly the high $\ell$ modes, especially for high signal to noise data \citep{Basak:2013}. These issues require specific attention to the exact response of the detectors, precise calibration of the instrument, especially re-calibration of data sets from different instruments inside the algorithm itself. In a follow up paper, \cite{Adak:2021} (In preparation), we demonstrate the application of cMILC algorithm on \wmap\ and \planck\ real data, re-calibration of the data in the same algorithm etc.   

Finally, this algorithm is in principle applicable to recover any foreground templates, moment maps of any order at any frequency. While we mainly focus on the estimation of foreground maps in the current paper, one can extend this work for cleaning the CMB \Qm, \Um maps from foreground contamination over incomplete sky. Furthermore, the moment expansion method is extremely useful and be applicable to extract the CMB spectral distortion signal \citep{Rotti:2020}, 21cm global signal, CMB B-mode signal \citep{Remazeilles:2020} etc. This approach also allows us to use external templates to minimise the contribution of extra components, a similar approach like the internal template fitting \citep{Fernandez-Cobos:2012}.      

\section*{Data Availability}
The \GAL\ mask is taken from PLA (\url{pla.esac.esa.int/pla/}).
\section*{Acknowledgements}

DA acknowledges the University Grants Commission India for providing financial support as Senior Research Fellow. This work was supported by Science and Engineering Research Board, Department of Science and Technology, Govt. of India grant number SERB/ECR/2018/000826. Some of the computations in this paper are done on the Pegasus cluster\footnote{\url{http://hpc.iucaa.in/}} at IUCAA. DA acknowledges Prof. Tarun Souradeep, Dr. Tuhin Ghosh and Dr. Shabbir Shaikh for useful discussion regarding this work.

\bibliographystyle{mn2e}
\bibliography{cmilc} 

\appendix

\section{cMILC results of the simulations in SET2 and SET3}
\label{sec:other_sim_results}
\begin{figure*}
\begin{multicols}{2}
    \includegraphics[width=\linewidth]{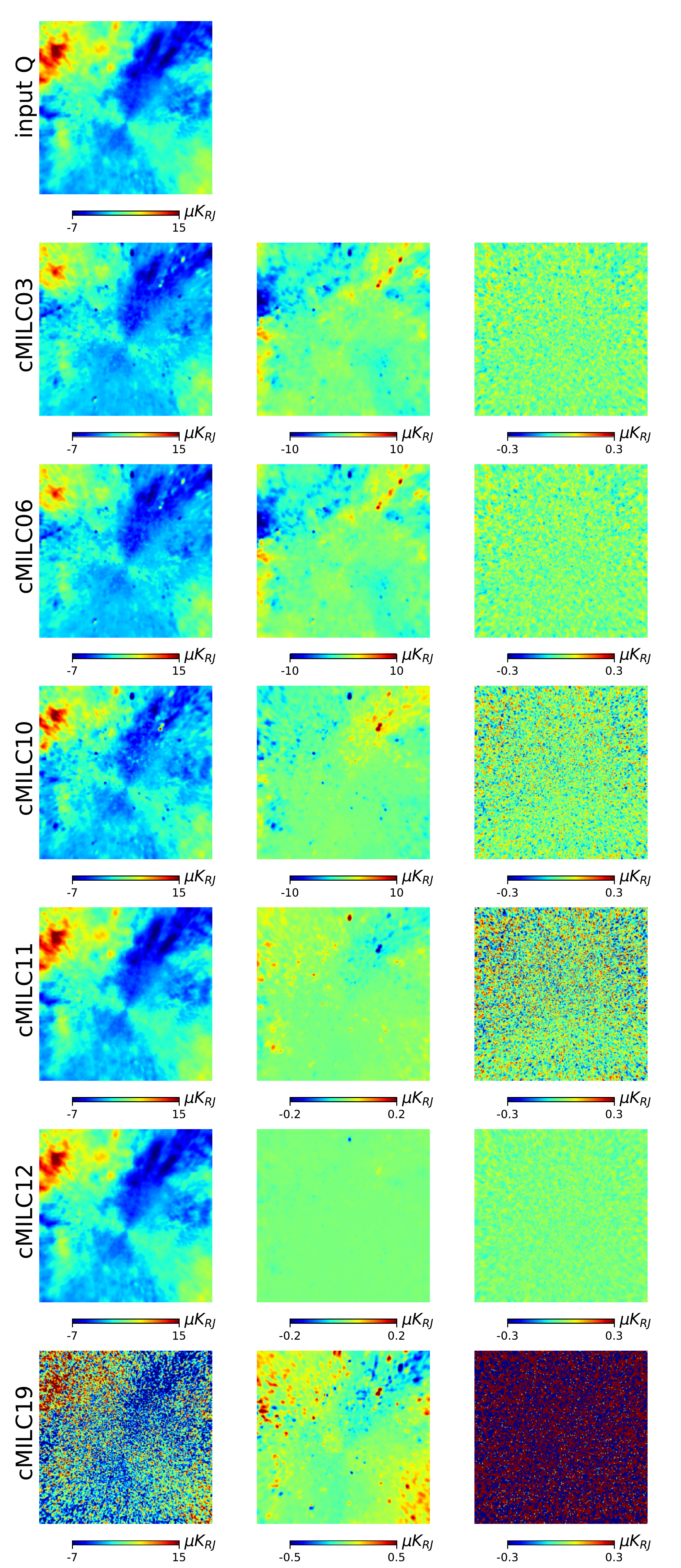} \par
    \includegraphics[width=\linewidth]{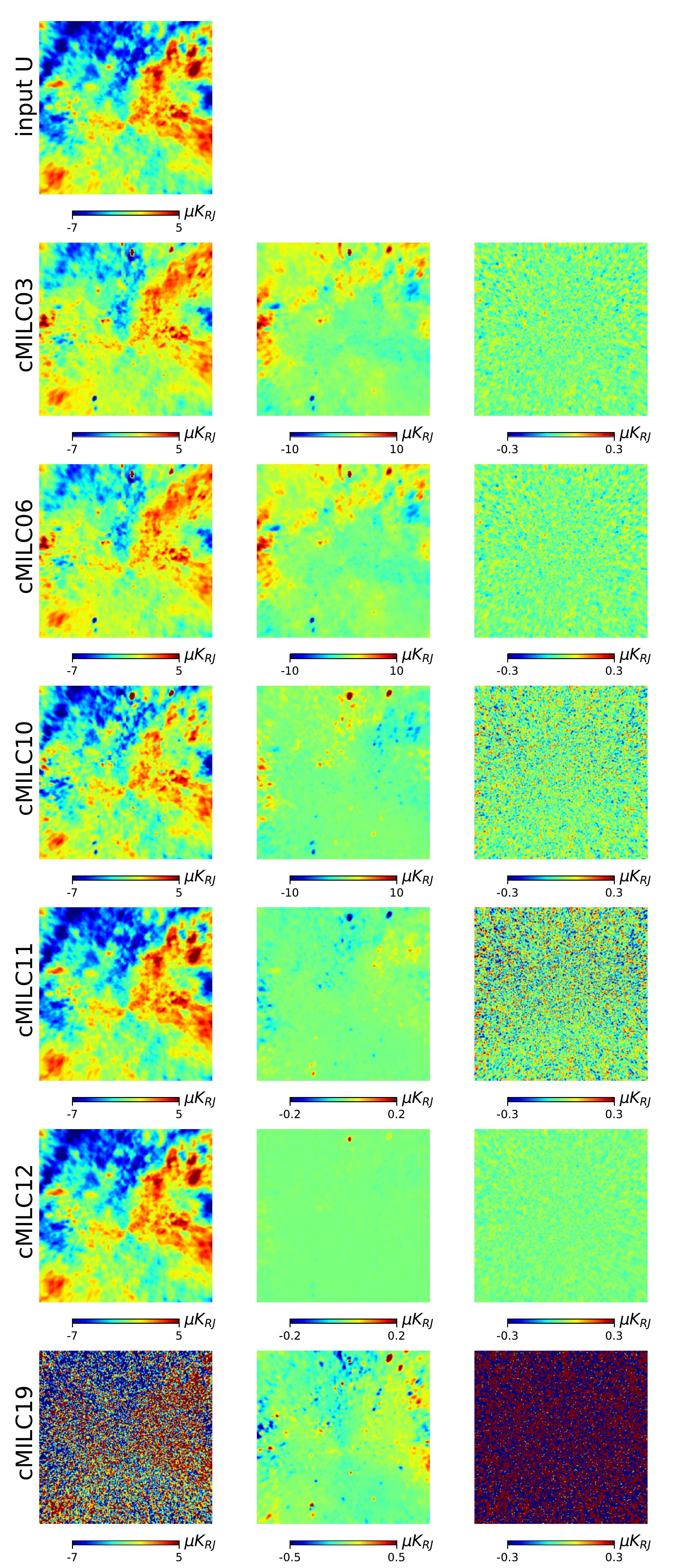} \par
    \end{multicols}
    \caption{\textit{Left panel} shows the cMILC results of estimation of thermal dust Q maps, and \textit{right panel} shows the cMILC results of estimation of thermal dust U maps for the simulation in SET2. The patches are 70$^{\deg}$ $\times$ 70$^{\deg}$ shown in gnomonic projection centered at $(l, b)$ = (90$^{\deg}$, -80$^{\deg}$). All maps are smoothed at a resolution of FWHM = 60\parcm.  The first rows show input thermal dust maps. The first, second and third columns of the subsequent rows show the recovered thermal dust maps, moment residual maps and noise residual maps for some selected cMILC iterations starting from cMILC03 to cMILC19.}
    \label{fig:dist_dust_maps_sim_d4s3}
\end{figure*}

\begin{figure*}
\begin{multicols}{2}
    \includegraphics[width=\linewidth]{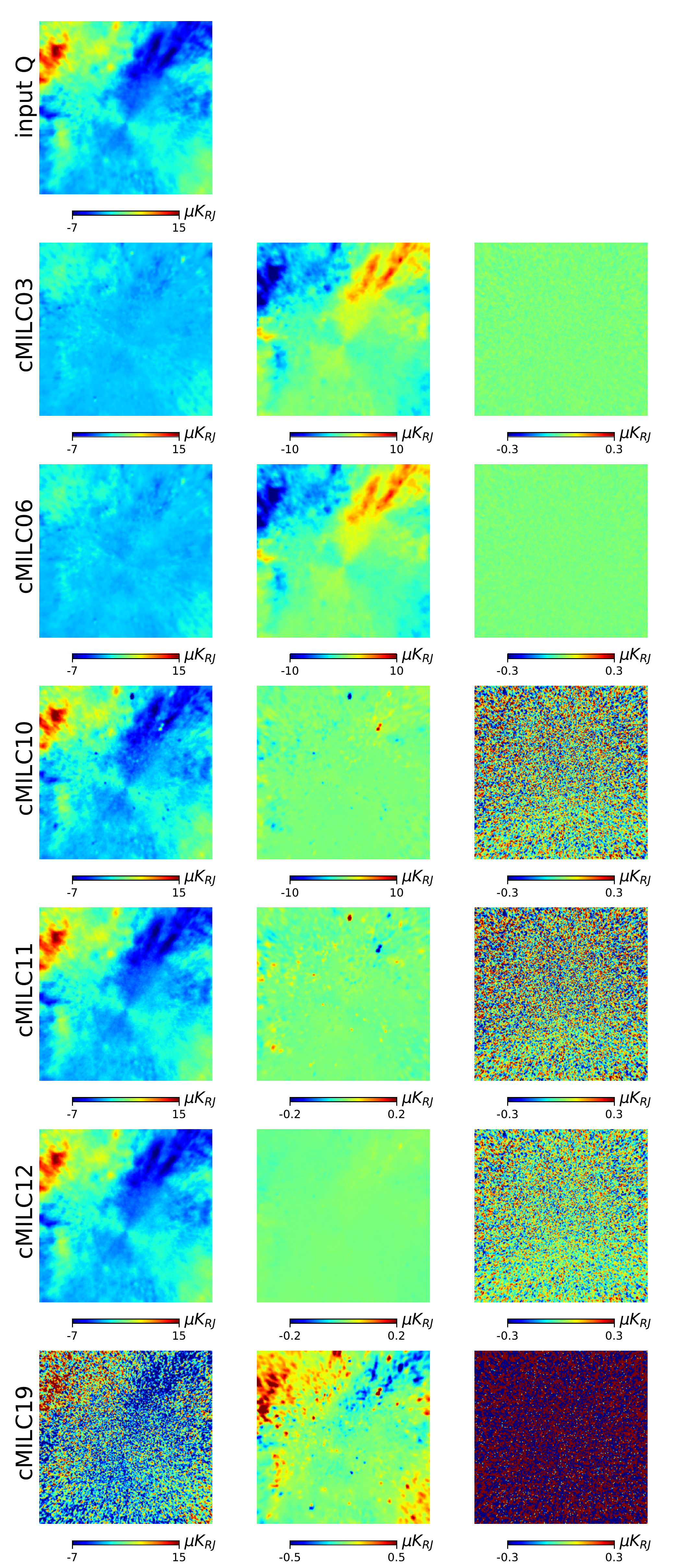} \par
    \includegraphics[width=\linewidth]{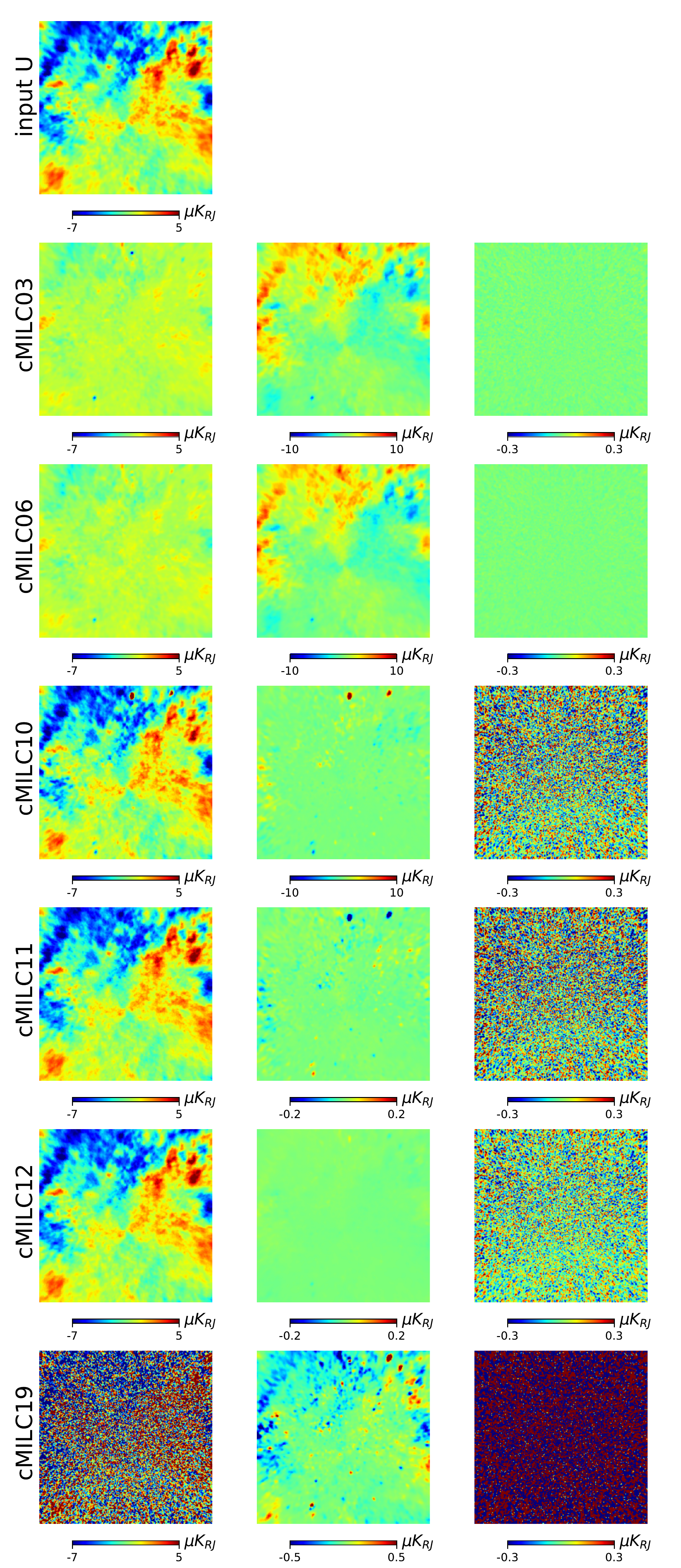} \par
    \end{multicols}
    \caption{\textit{Left panel} shows the cMILC results of estimation of thermal dust Q maps, and \textit{right panel} shows the cMILC results of estimation of thermal dust U maps for the simulation in SET3. The patches are 70$^{\deg}$ $\times$ 70$^{\deg}$ shown in gnomonic projection centered at $(l, b)$ = (90$^{\deg}$, -80$^{\deg}$). All maps are smoothed at a resolution of FWHM = 60\parcm.  The first rows show input thermal dust maps. The first, second and third columns of the subsequent rows show the recovered thermal dust maps, moment residual maps and noise residual maps for some selected cMILC iterations starting from cMILC03 to cMILC19.}
    \label{fig:dist_dust_maps_sim_d7s2}
\end{figure*}

\begin{figure*}
\begin{multicols}{2}
    \includegraphics[width=\linewidth]{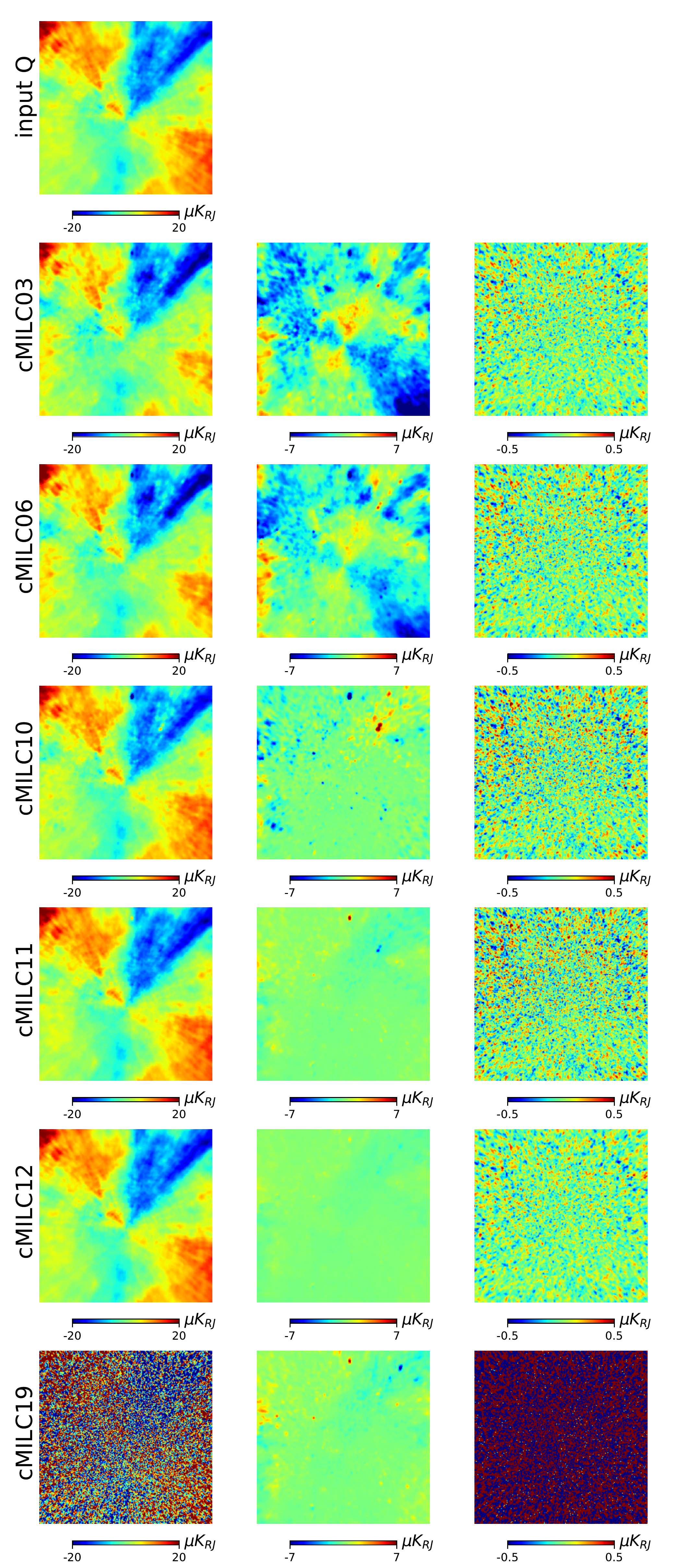} \par
    \includegraphics[width=\linewidth]{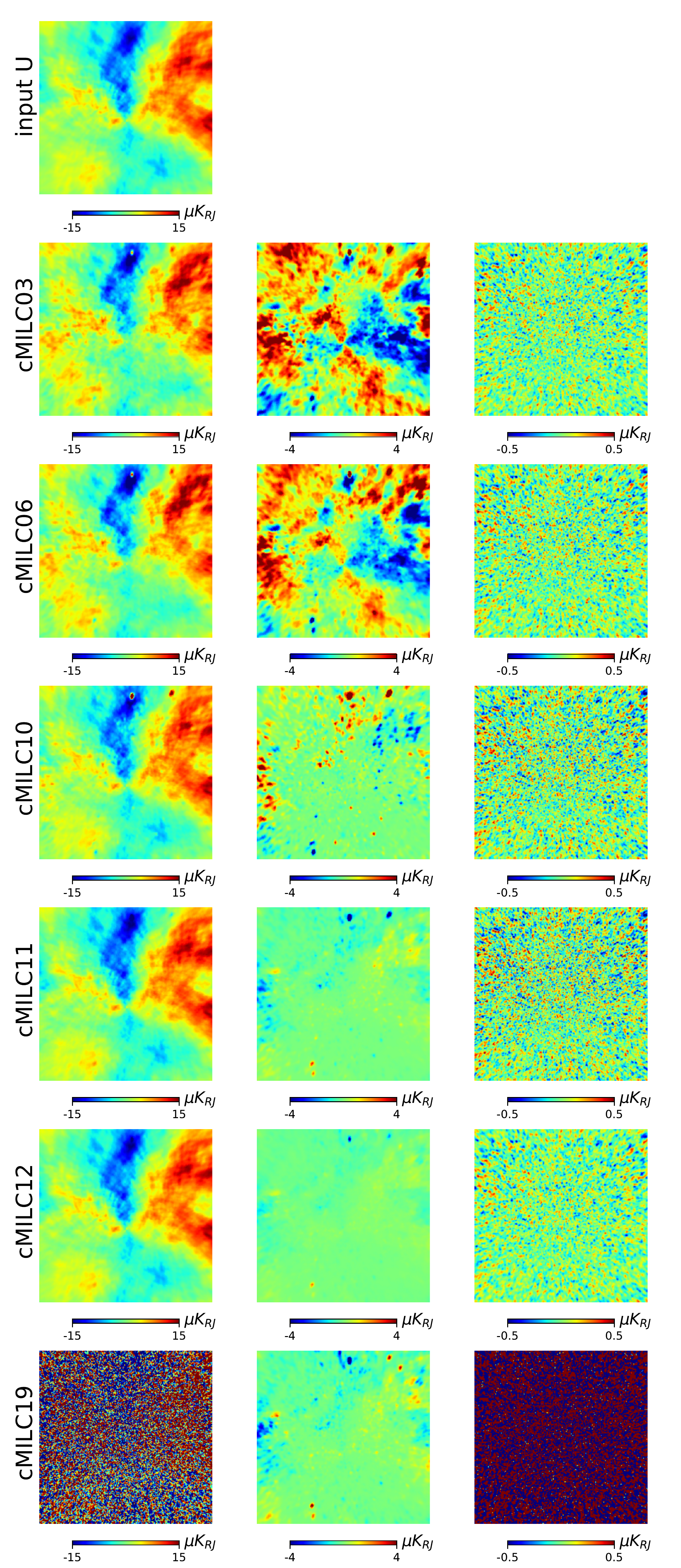} \par
    \end{multicols}
    \caption{\textit{Left panel} shows the cMILC results of estimation of synchrotron Q maps, and \textit{right panel} shows the cMILC results of estimation of synchrotron U maps for the simulation in SET2. The patches are 70$^{\deg}$ $\times$ 70$^{\deg}$ shown in gnomonic projection centered at $(l, b)$ = (90$^{\deg}$, -80$^{\deg}$). All maps are smoothed at a resolution of FWHM = 60\parcm.  The first rows show input synchrotron maps. The first, second and third columns of the subsequent rows show the recovered synchrotron maps, moment residual maps and noise residual maps for some selected cMILC iterations starting from cMILC03 to cMILC19.}
    \label{fig:dist_sync_maps_sim_d4s3}
\end{figure*}

\begin{figure*}
\begin{multicols}{2}
    \includegraphics[width=\linewidth]{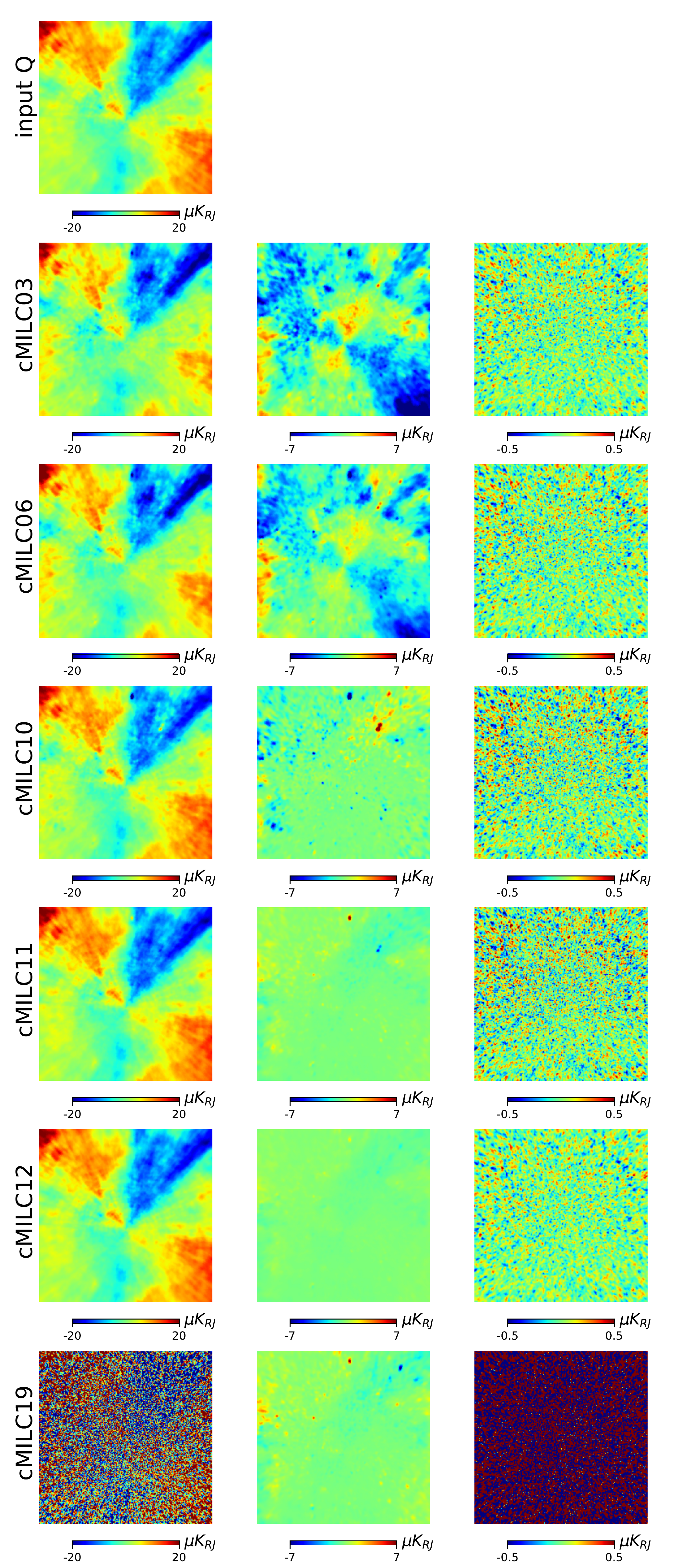} \par
    \includegraphics[width=\linewidth]{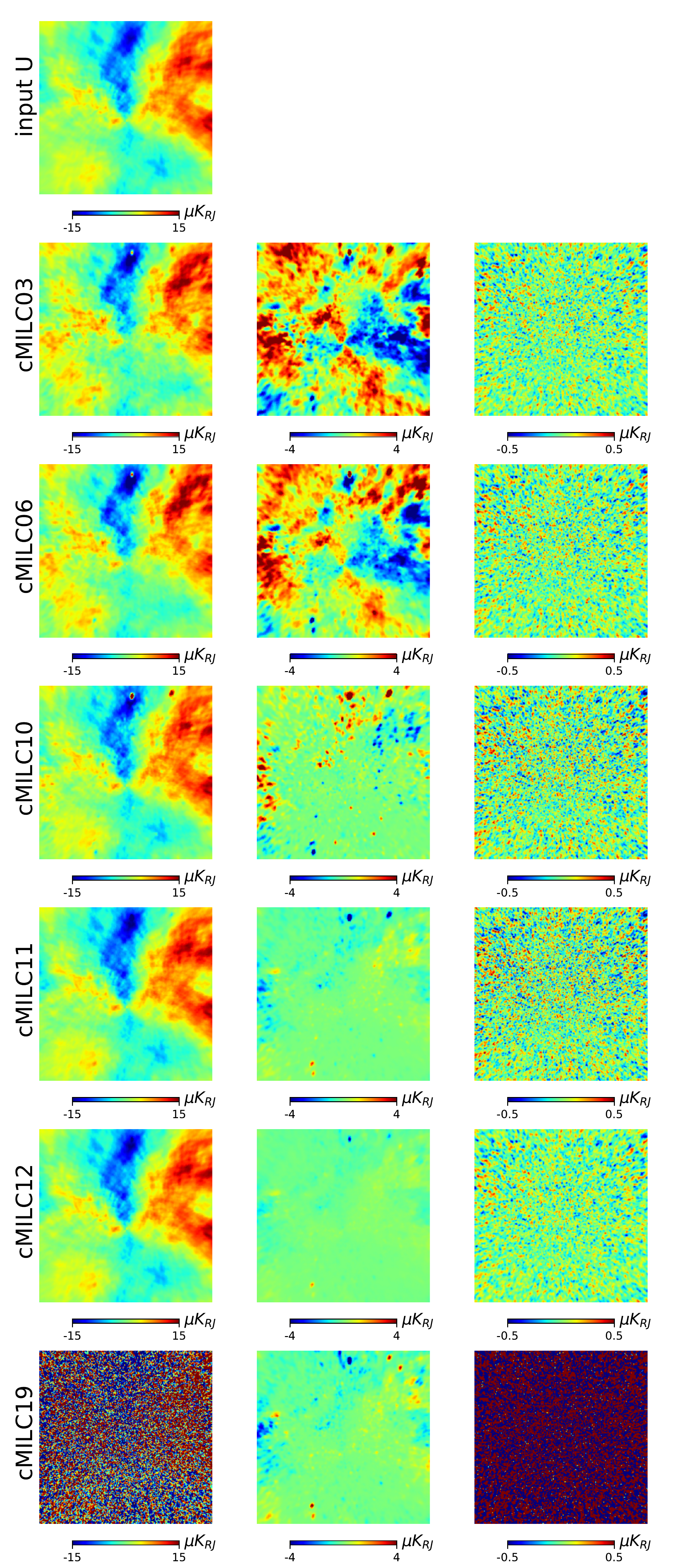} \par
    \end{multicols}
    \caption{\textit{Left panel} shows the cMILC results of estimation of synchrotron Q maps, and \textit{right panel} shows the cMILC results of estimation of synchrotron U maps for the simulation in SET3. The patches are 70$^{\deg}$ $\times$ 70$^{\deg}$ shown in gnomonic projection centered at $(l, b)$ = (90$^{\deg}$, -80$^{\deg}$). All maps are smoothed at a resolution of FWHM = 60\parcm.  The first rows show input synchrotron maps. The first, second and third columns of the subsequent rows show the recovered synchrotron maps, moment residual maps and noise residual maps for some selected cMILC iterations starting from cMILC03 to cMILC19.}
    \label{fig:dist_sync_maps_sim_d7s2}
\end{figure*}

\begin{figure*}
    
    \begin{multicols}{2}
    \includegraphics[width=\linewidth]{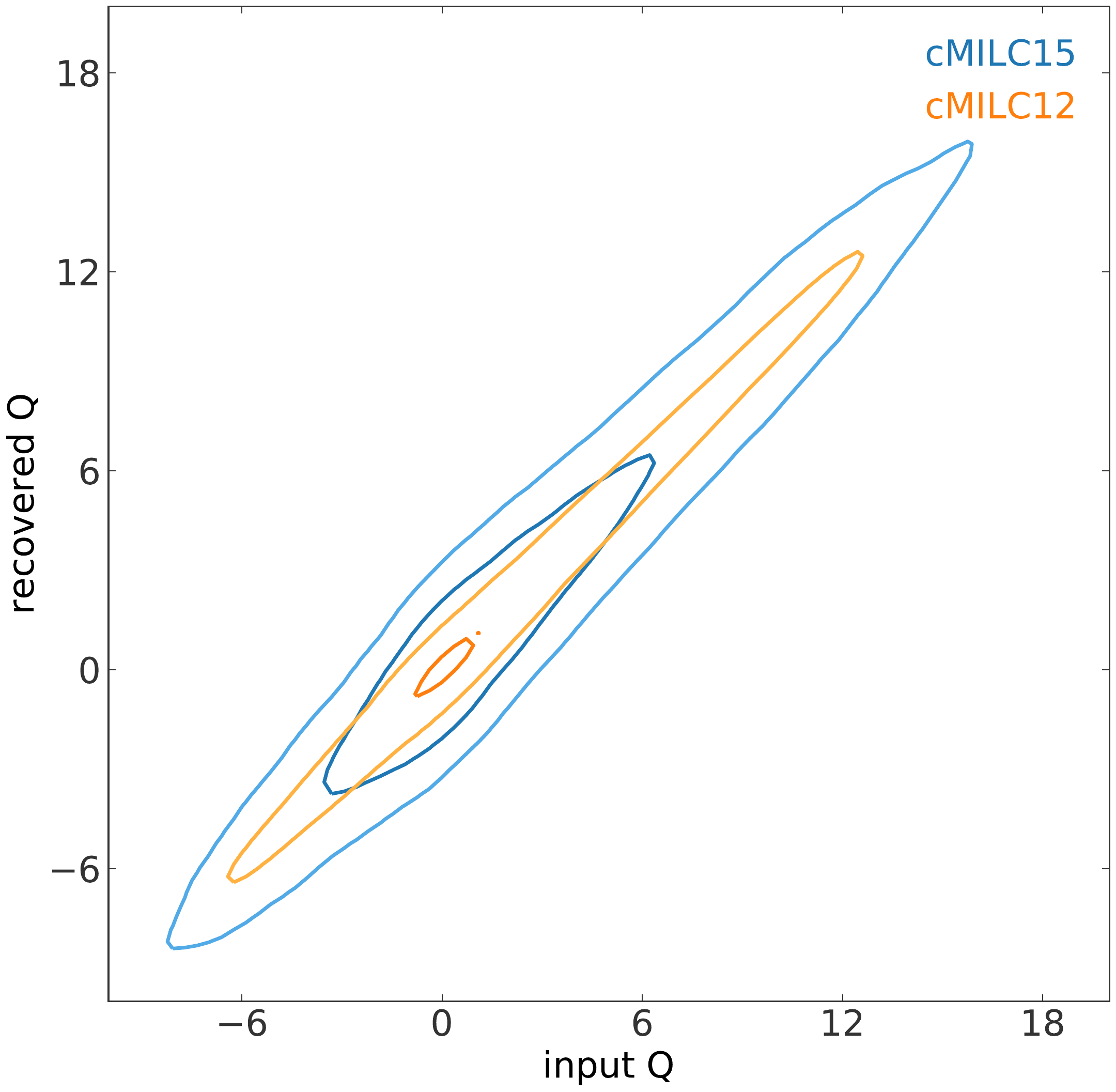}\par
    \includegraphics[width=\linewidth]{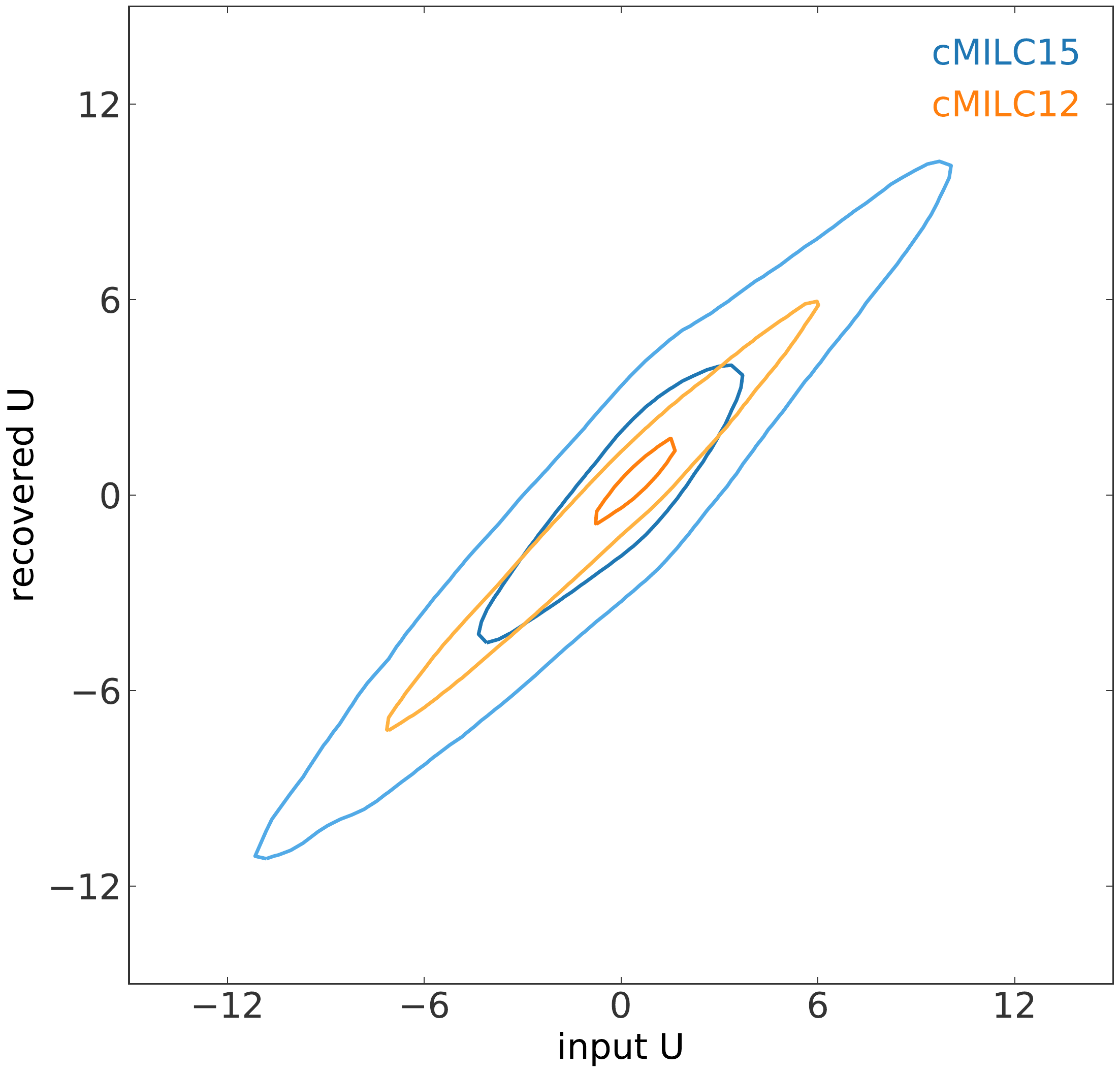}\par
    \end{multicols}
    \begin{multicols}{2}
    \includegraphics[width=\linewidth]{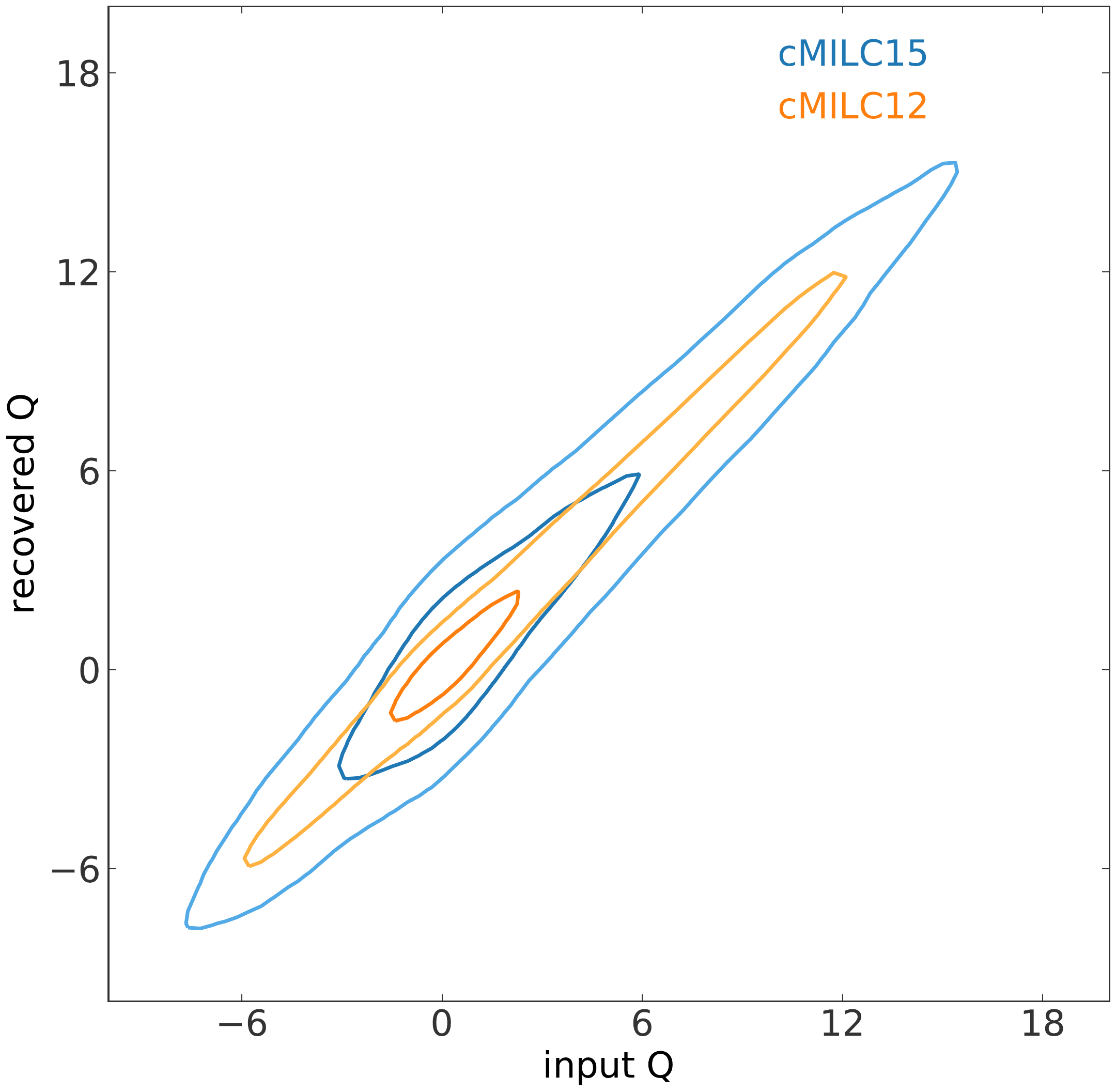}\par
    \includegraphics[width=\linewidth]{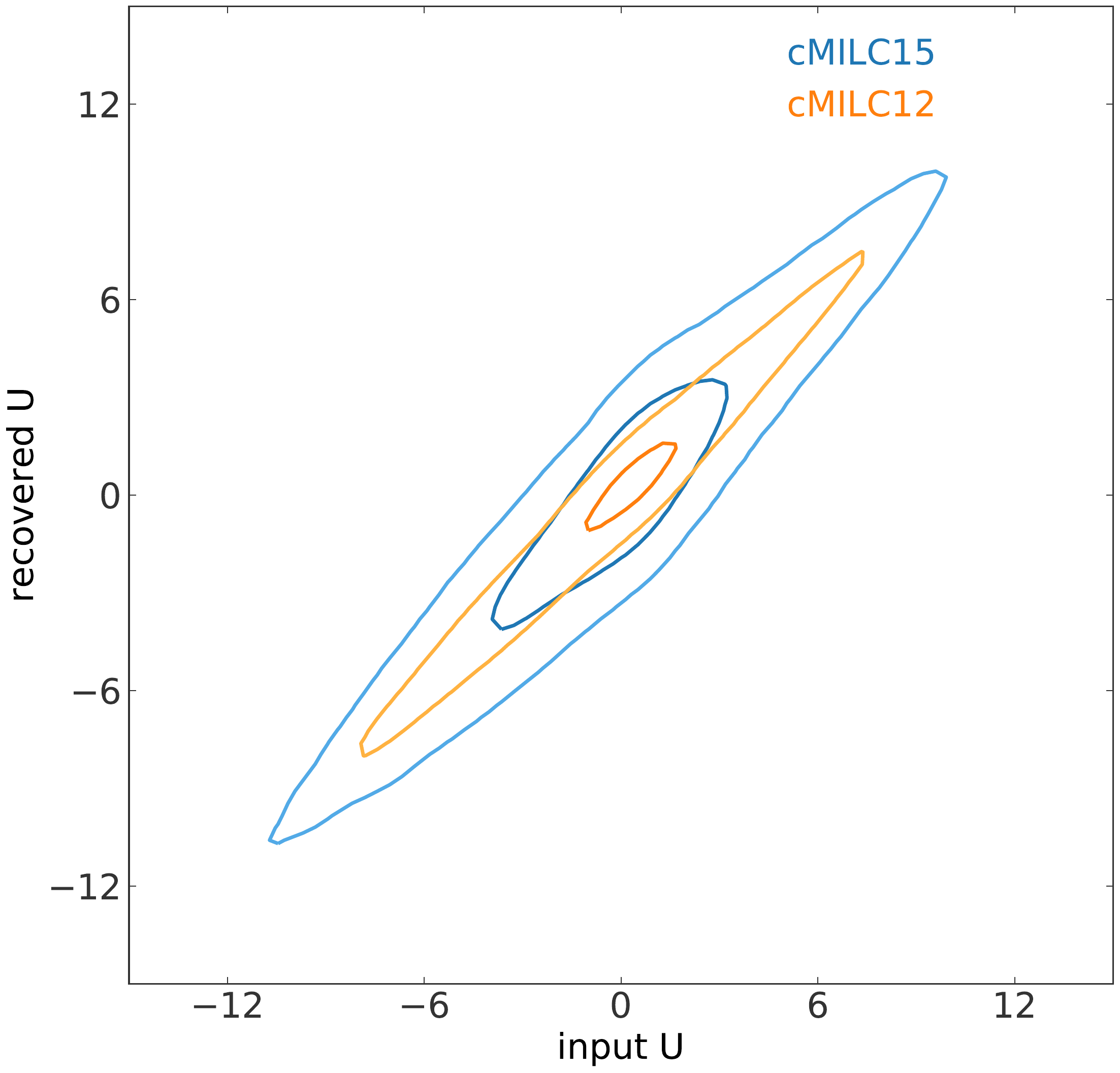}\par
    \end{multicols}
    \caption{Contour plots of 2D-histogram of input \Qm (\textit{left panel}) and \Um (\textit{right panel}) thermal dust maps and recovered thermal dust maps. 1$\sigma$ and 2$\sigma$ contours are shown here for cMILC12 (orange) and cMILC15 (blue) iterations. Results for simulation in SET2 is presented in \textit{upper panel} and results for simulation in SET3 is presented in \textit{lower panel}.}
    \label{fig:dust_TT_correlation_d4s3_d7s2}
\end{figure*}

\begin{figure*}
    \begin{multicols}{2}
    \includegraphics[width=\linewidth]{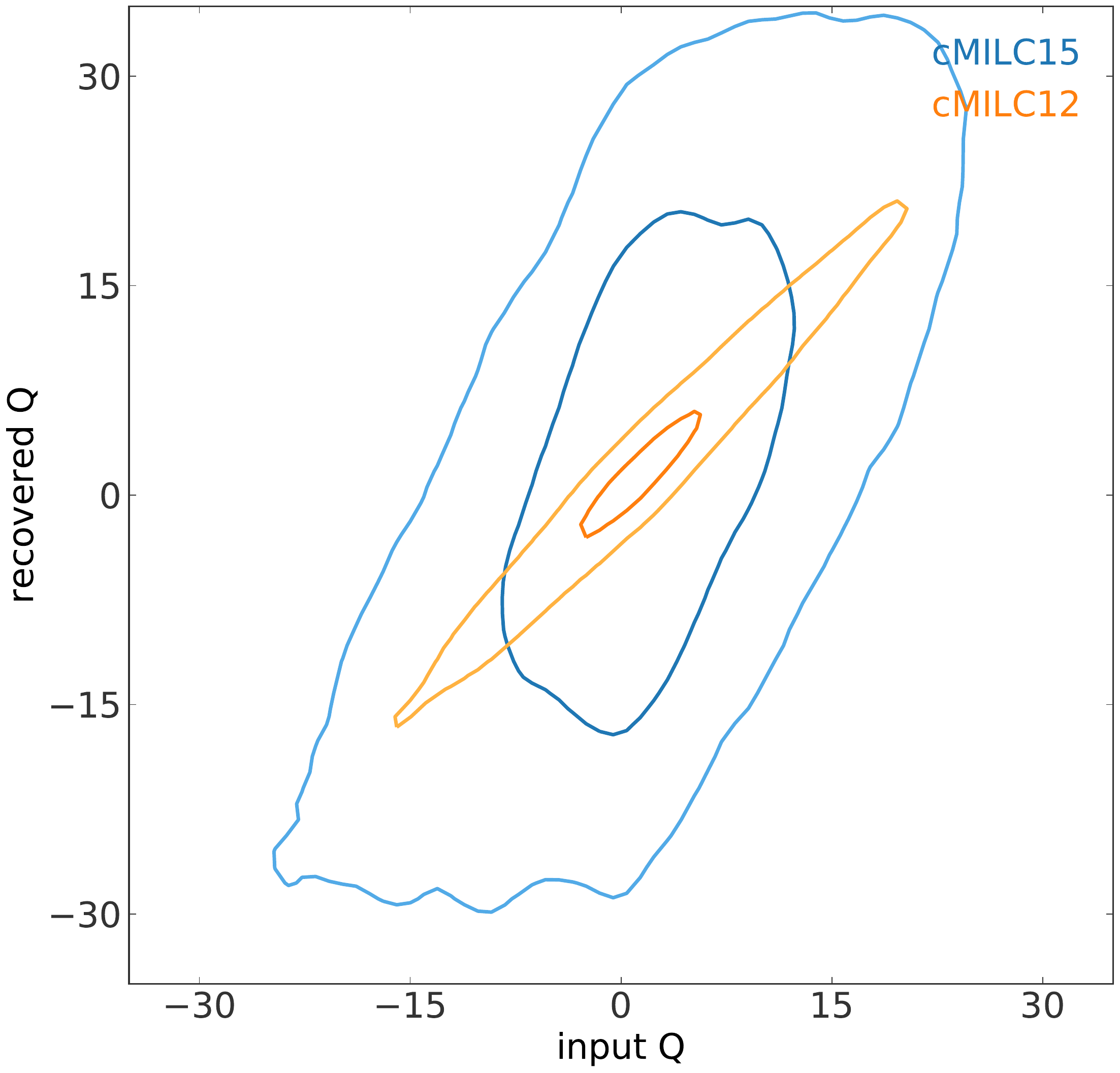}\par
    \includegraphics[width=\linewidth]{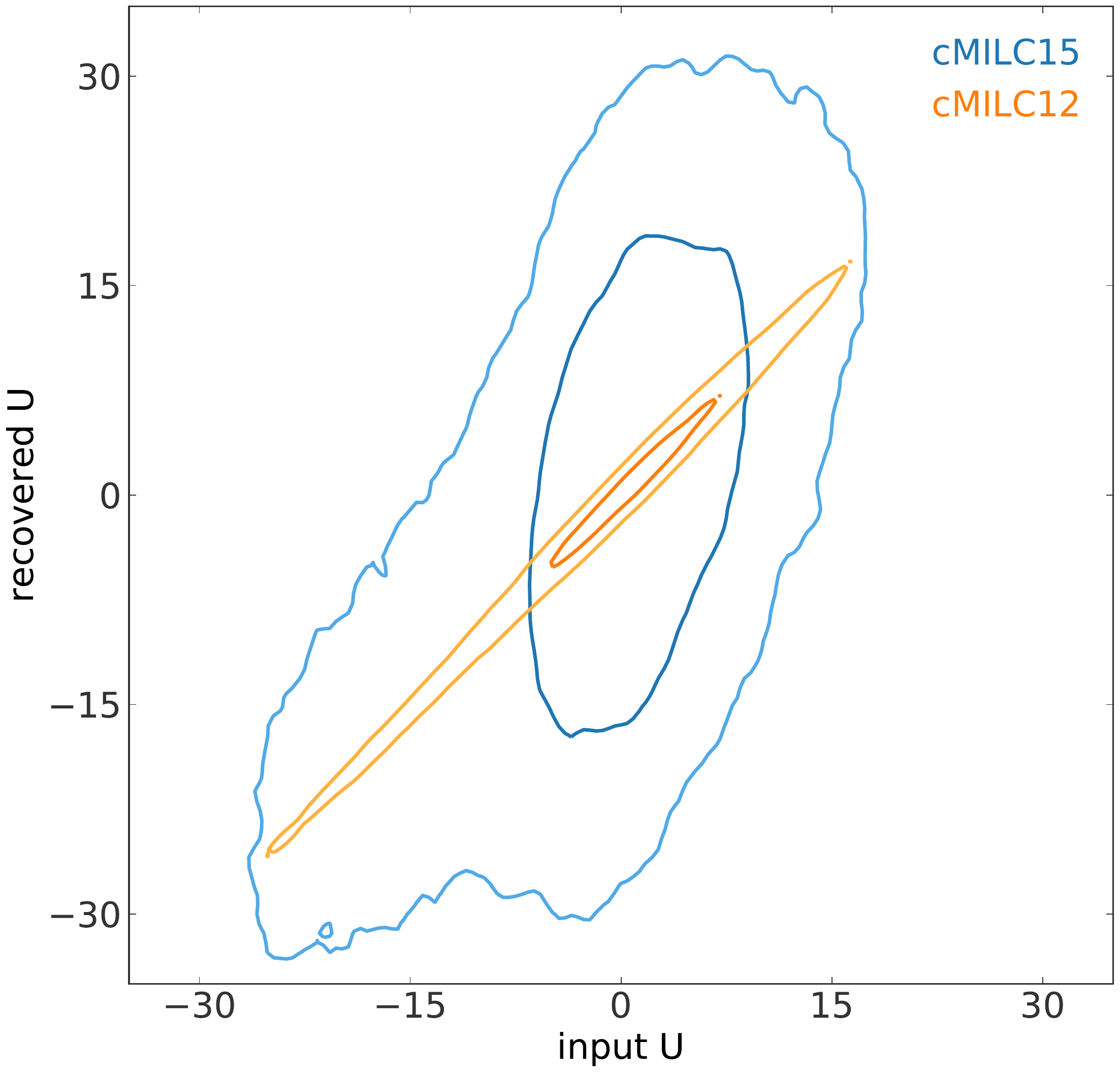}\par
    \end{multicols}
    \begin{multicols}{2}
    \includegraphics[width=\linewidth]{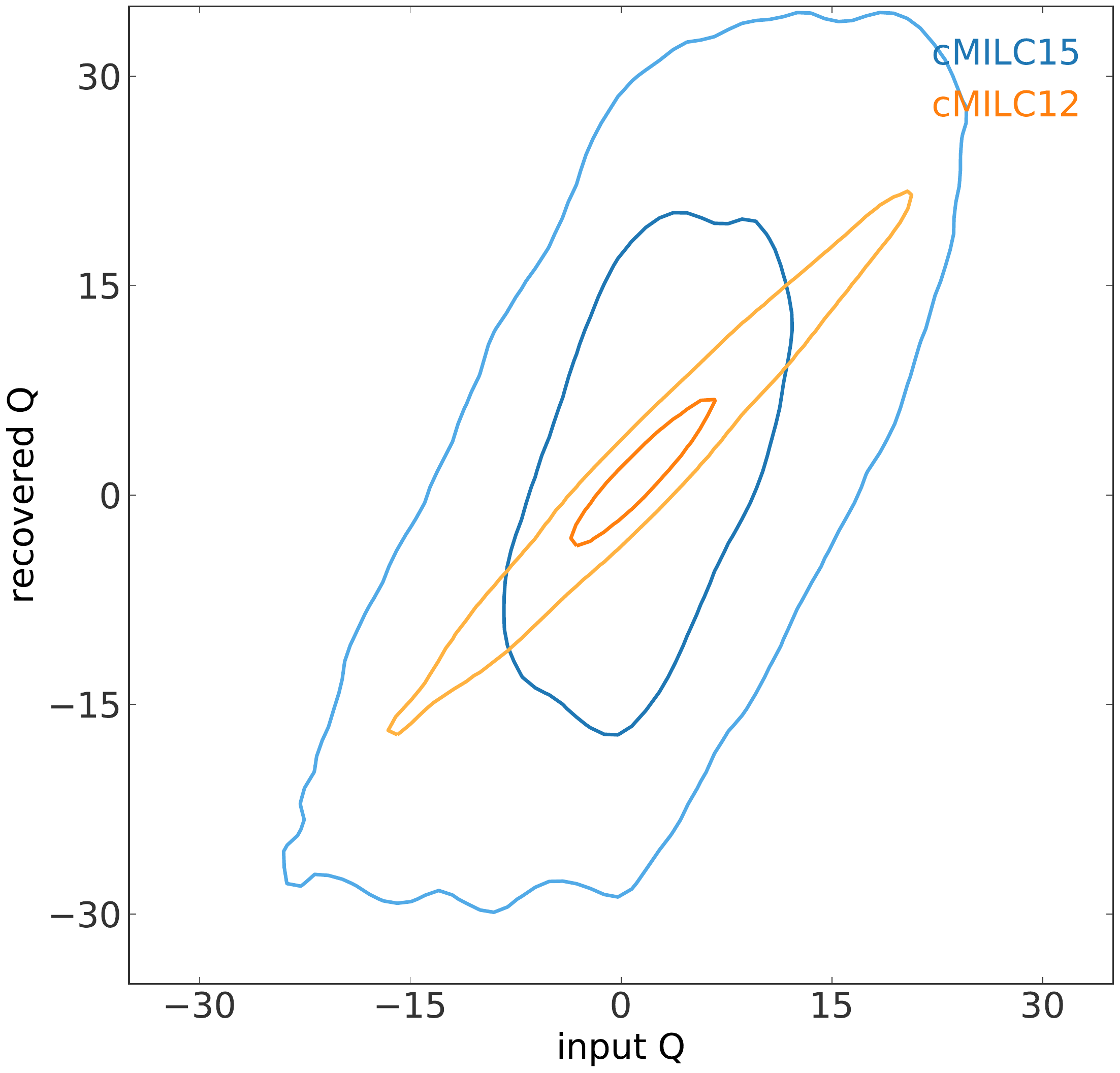}\par
    \includegraphics[width=\linewidth]{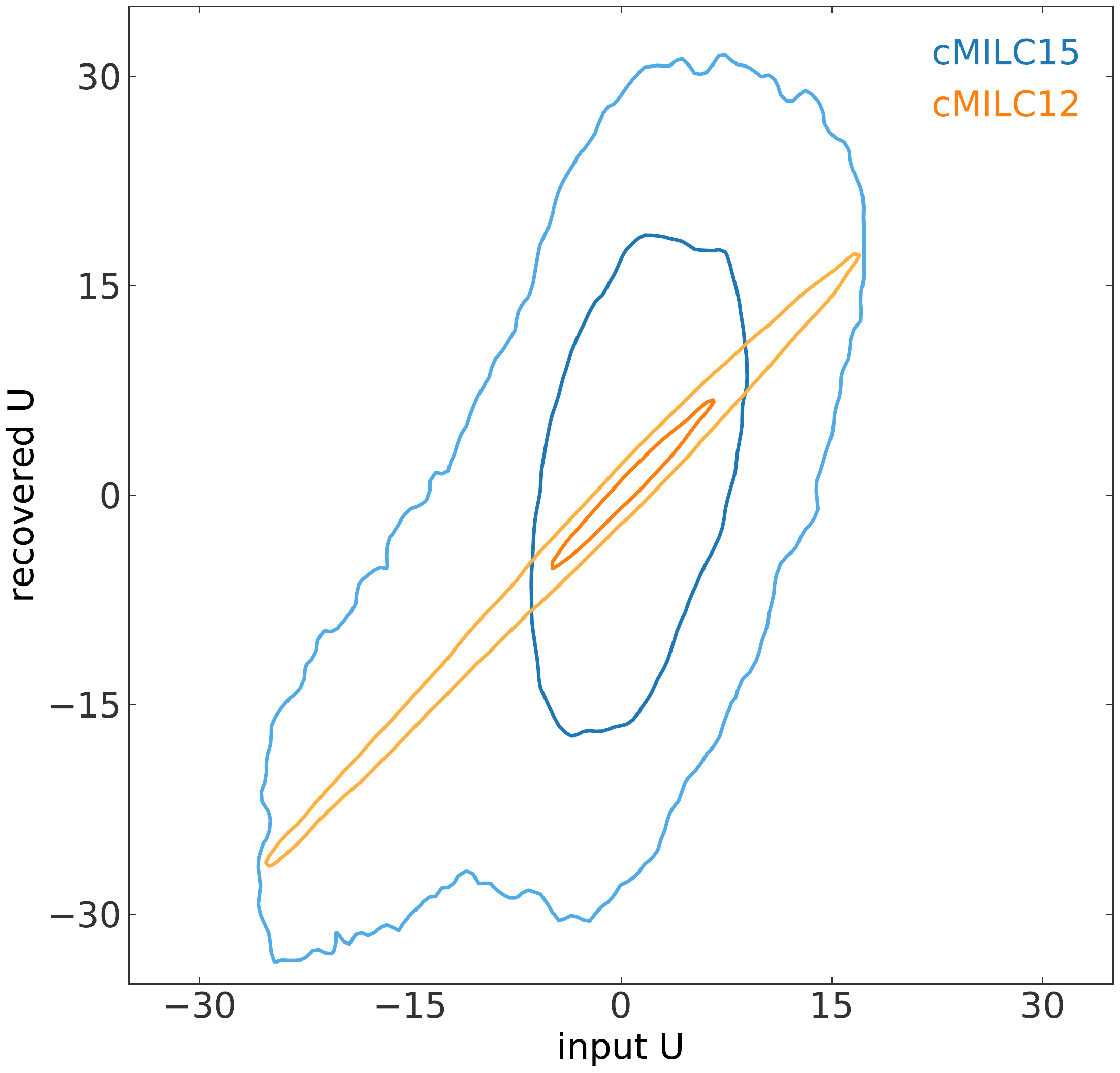}\par
    \end{multicols}
    \caption{Contour plots of 2D-histogram of input \Qm (\textit{left panel}) and \Um (\textit{right panel}) synchrotron maps and recovered synchrotron maps. 1$\sigma$ and 2$\sigma$ contours are shown here for cMILC12 (orange) and cMILC15 (blue) iterations. Results for simulation in SET2 is presented in \textit{upper panel} and results for simulation in SET3 is presented in \textit{lower panel}.}
    \label{fig:sync_TT_correlation_d4s3_d7s2}
\end{figure*}

\begin{figure*}
    \begin{multicols}{2}
    \includegraphics[width=\linewidth]{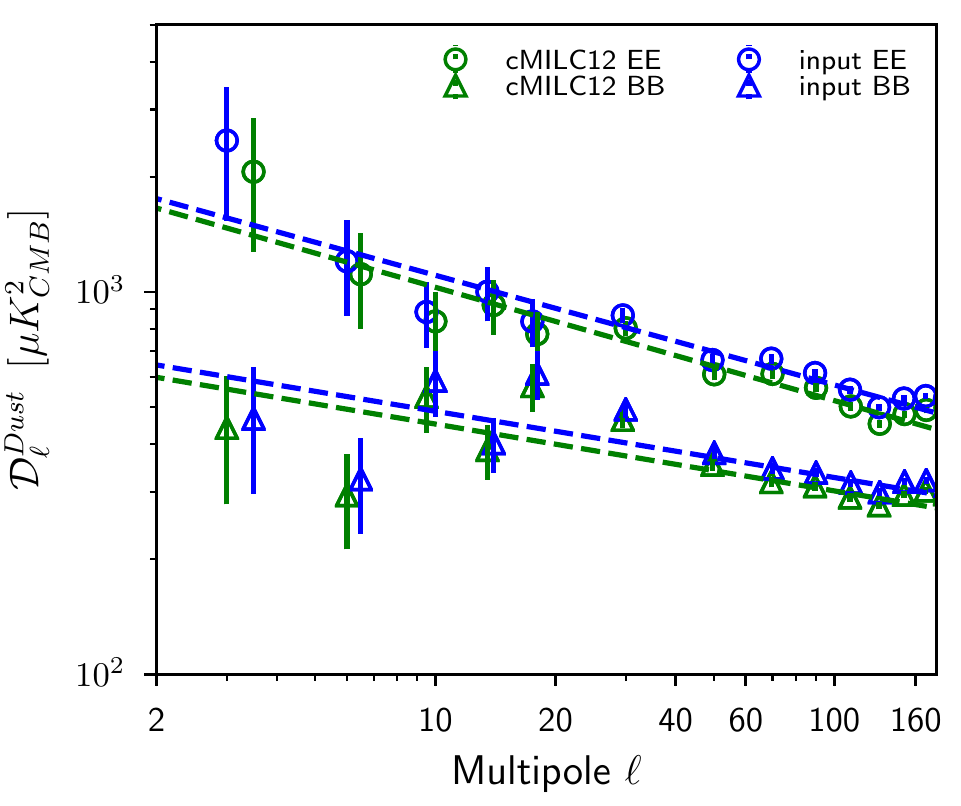} \par
    \includegraphics[width=\linewidth]{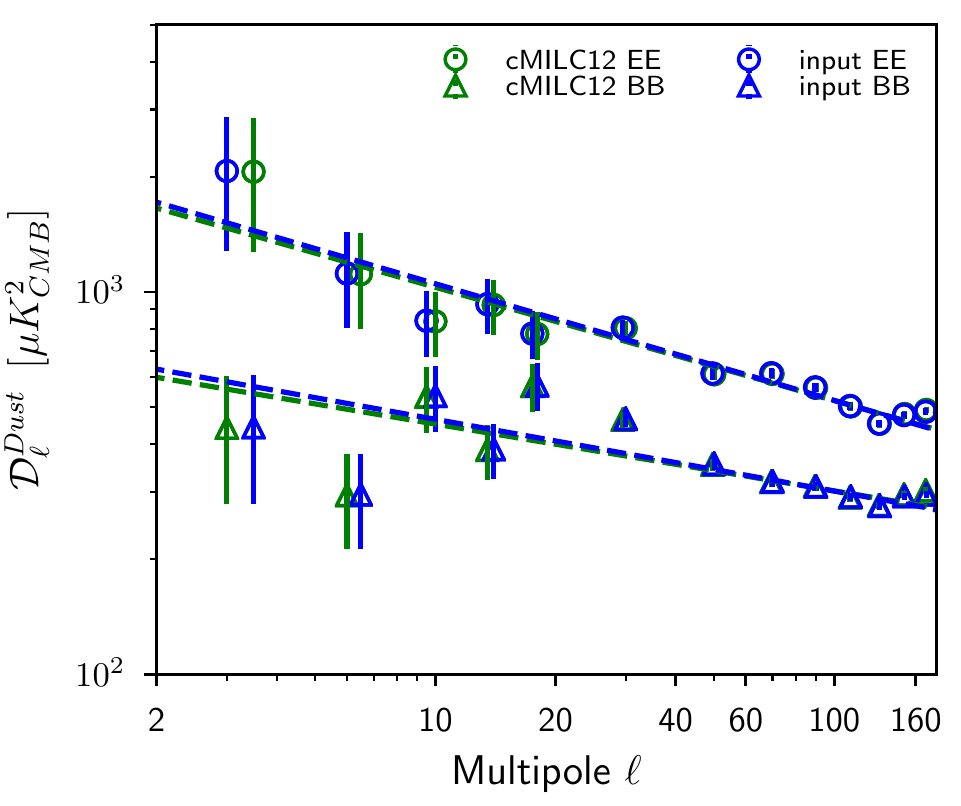} \par
    \end{multicols}
    \caption{EE (circles) and BB (triangles) power spectra of thermal dust maps for simulation in SET2 (\textit{left panel}) and SET3 (\textit{right panel}). Spectra estimated from the input maps are shown in blue, and spectra estimated from the recovered maps for cMILC12 iteration are shown in green. All spectra are computed over \GAL\ apodized mask using \xpol. Error bars are 1$\sigma$ uncertainties analytically computed from \xpol.}
    \label{fig:sim_dust_power_d4s3_d7s2}
\end{figure*}

\begin{figure*}
    \begin{multicols}{2}
    \includegraphics[width=\linewidth]{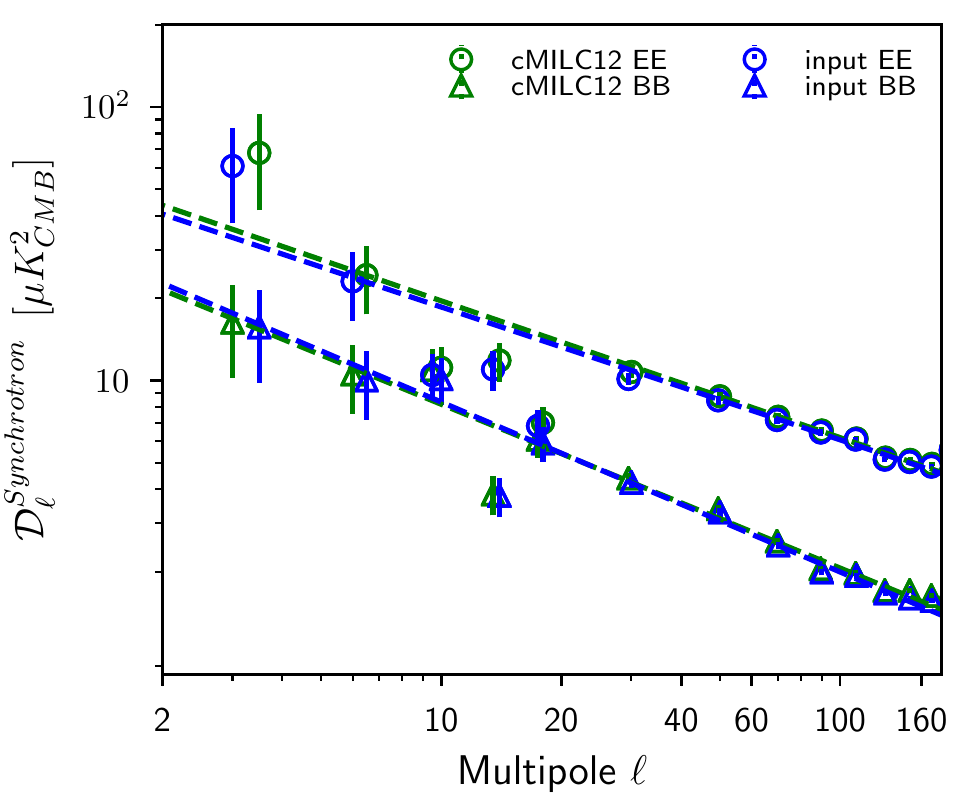} \par
     \includegraphics[width=\linewidth]{Sync_power_spectra_data_modeld7s2a2_maps_ns256_cMILC12.pdf} \par
    \end{multicols}
    \caption{EE (circles) and BB (triangles) power spectra of synchrotron maps for simulation in SET2 (\textit{left panel}) and SET3 (\textit{right panel}). Spectra estimated from the input maps are shown in blue, and spectra estimated from the recovered maps for cMILC12 iteration are shown in green. All spectra are computed over  \GAL\ apodized mask using \xpol. Error bars are 1$\sigma$ uncertainties analytically computed from \xpol.}
    \label{fig:sim_sync_power_d4s3_d7s2}
\end{figure*}

In this section, we present the same results for simulations in SET2 and SET3 as presented in main text for SET1. In Figure.~\ref{fig:dist_dust_maps_sim_d4s3} and Figure.~\ref{fig:dist_dust_maps_sim_d7s2}, we respectively summarize the results of estimation of thermal dust \Qm (\textit{left panel}) and \Um (\textit{right panel}) templates for simulation in SET2 and SET3 for some selected cMILC iterations with increasing constraints. 70$^{\deg}$ $\times$ 70$^{\deg}$ patches are displayed in gnomonic projection centered at the Galactic longitude and latitude, $(l, b)$ = (90$^{\deg}$, -80$^{\deg}$). Similarly, In Figure.~\ref{fig:dist_sync_maps_sim_d4s3} and Figure.~\ref{fig:dist_sync_maps_sim_d7s2}, we respectively summarize the similar results of estimation of synchrotron template for simulation in SET2 and SET3 for some selected cMILC iterations. The same patches are show in in gnomonic projection. In Figure.~\ref{fig:dust_TT_correlation_d4s3_d7s2}, we display the 1$\sigma$ and 2$\sigma$ contours of 2D-histogram of input thermal dust and recovered thermal dust maps for two selected cMILC iterations (cMILC12 and cMILC15). Results for \Qm and \Um are shown in \textit{left panel} and \textit{right panel} respectively. \textit{Upper panel} shows results of the simulation in SET2 and \textit{lower panel} shows the results of the simulation in SET3. Similar results for synchrotron are presented in Figure.~\ref{fig:sync_TT_correlation_d4s3_d7s2}. We compare input and estimated thermal dust power spectra for cMILC12 in Figure.~\ref{fig:sim_dust_power_d4s3_d7s2}. \textit{Left panel} and \textit{right panel} shows the results for SET2 and SET3 respectively. Same comparison for synchrotron is shown in Figure.~\ref{fig:sim_sync_power_d4s3_d7s2}.

\end{document}